\documentclass{IEEEtran}
\usepackage{cite}
\usepackage{amsmath,amssymb,amsfonts}
\usepackage{algorithmic}
\usepackage{graphicx}
\usepackage{textcomp}
\usepackage{subfigure}
\usepackage{makecell}
\usepackage{color}
\usepackage{bm}
\usepackage{algorithm,algorithmic}
\def\BibTeX{{\rm B\kern-.05em{\sc i\kern-.025em b}\kern-.08em
    T\kern-.1667em\lower.7ex\hbox{E}\kern-.125emX}}
\begin{document}
\title{Multichannel Synthetic Preictal EEG Signals to Enhance the Prediction of Epileptic Seizures}
\author{Yankun Xu, Jie Yang, and Mohamad Sawan, \IEEEmembership{Fellow, IEEE}
\thanks{Yankun Xu is with the Zhejiang University, Hangzhou, Zhejiang, China and with the Center of Excellence in Biomedical Research on Advanced Integrated-on-chips Neurotechnologies (CenBRAIN Neurotech), School of Engineering, Westlake University, Hangzhou, Zhejiang, China.}
\thanks{Jie Yang and Mohamad Sawan are with the Center of Excellence in Biomedical Research on Advanced Integrated-on-chips Neurotechnologies (CenBRAIN Neurotech), School of Engineering, Westlake University, Hangzhou, Zhejiang, China, (e-mail: yangjie@westlake.edu.cn).}}

\maketitle

\begin{abstract}
Epilepsy is a chronic neurological disorder affecting 1\% of people worldwide, deep learning (DL) algorithms-based electroencephalograph (EEG) analysis provides the possibility for accurate epileptic seizure (ES) prediction, thereby benefiting patients suffering from epilepsy. To identify the preictal region that precedes the onset of seizure, a large number of annotated EEG signals are required to train DL algorithms. However, the scarcity of seizure onsets leads to significant insufficiency of data for training the DL algorithms. To overcome this data insufficiency, in this paper, we propose a preictal artificial signal synthesis algorithm based on a generative adversarial network to generate synthetic multichannel EEG preictal samples. A high-quality single-channel architecture, determined by visual and statistical evaluations, is used to train the generators of multichannel samples. The effectiveness of the synthetic samples is evaluated by comparing the ES prediction performances without and with synthetic preictal sample augmentation. The leave-one-seizure-out cross validation ES prediction accuracy and corresponding area under the receiver operating characteristic curve evaluation improve from 73.0\% and 0.676 to 78.0\% and 0.704 by 10$\times$ synthetic sample augmentation, respectively. The obtained results indicate that synthetic preictal samples are effective for enhancing ES prediction performance.
\end{abstract}

\begin{IEEEkeywords}
Epilepsy, preictal, seizure prediction, machine learning, deep learning, generative adversarial network, data augmentation, synthetic EEG
\end{IEEEkeywords}

\section{Introduction}
\label{sec:introduction}
Epilepsy is one of the most common chronic neurodegenerative disorders characterized by recurrent epileptic seizures (ESs) triggered by abnormal discharge of neurons \cite{world2006neurological,yang2020seizure}. Approximately 1\% people worldwide suffer from unprovoked ESs, and the primary treatment for epileptic patients is long-term drug therapy. Moreover, approximately one-third of patients are resistant to anti-epileptic medication \cite{assi2017towards,xu2021trends}. Epileptic patients can benefit from an accurate prediction system that enables early warning minutes before onset. This allows for precautional intervention or therapies, thereby alleviating the psychological burden on patients and the demand for healthcare resources.

\begin{figure}[!t]
\centerline{\includegraphics[width=\columnwidth]{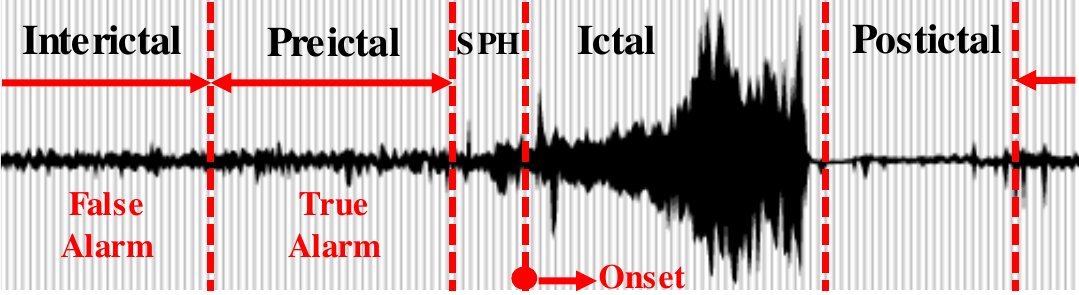}}
\caption{Example of an EEG illustrating different periods during a seizure occurrence.}
\label{fig1}
\end{figure}

Electroencephalography (EEG) is a prevalent neuroimaging approach widely used for epilepsy diagnosis and analysis. Recent research activities in the field of ES prediction are mainly based on either scalp EEG or invasive EEG signals \cite{rasheed2020machine,kuhlmann2018seizure,hu2020epileptic,1198251}. Fig. \ref{fig1} shows an annotated example of a single-channel EEG recording during a seizure occurrence. The recording of epileptic patients can be categorized into several periods: ictal (seizure occurrence), preictal (before seizure onset), postictal (shortly after the seizure), seizure prediction horizon (SPH, a short period before seizure onset in which precautional intervention is required) and interictal (between seizures excluding all other periods).

Over the past decades, various machine learning (ML) and deep learning (DL) based algorithms have been proposed to improve the performance of ES prediction \cite{7365453,7501827,8239676,zhao2020binary,5415597}. To train a reliable classifier, either ML- or DL-based algorithms requires a large number of annotated data. However, the unavailability of sufficient annotated preictal samples is a major challenge for ES prediction studies. First, the acquisition of sufficient annotated seizure data is costly and time-consuming. Second, the preictal period is only observed if a seizure occurs. Third, only one or a few seizures usually occur per day for most patients, in extreme cases that patients with intractable epilepsy can suffer from multiple seizures per day. Therefore the duration of interictal periods is significantly longer than that of preictal periods. The natural difference between interictal and preictal durations leads to the numbers of the two types of samples for the training set to be significantly imbalanced.

Recently, several researchers have applied generative adversarial network (GAN) technique to ES studies \cite{hartmann2018eeg,rasheed2020generative,truong2019epileptic}. However, they focus on the synthesis of either short-term ($<$5 s) single-channel raw EEG ictal signals or specific features extracted from raw EEG signals. These pioneering efforts cannot be used to synthesize long-term ($>5$ s) multichannel signals. This is primarily because this type of signals easily lead to mode collapse of GAN training. Additionally, the extracted features of preictal and ictal signals are significantly different even when they appear similar. Furthermore, the lack of an evaluation method for synthetic time-series data is always a challenge for GAN studies \cite{esteban2017real}.

In this paper, we propose a data generation method based on GAN to deliver synthetic long-duration multichannel time-series EEG preictal samples to address the problem of scarcity of preictal data and imbalanced training sets. The main contributions of this study are as follows:
\begin{itemize}
\item Generative model architectures based on convolutional neural network (CNN) and recurrent neural network (RNN) have been proposed to generate synthetic single-channel EEG preictal samples. Dedicated statistical evaluation metrics are also proposed to assess the quality of single-channel samples in order to choose the outperformed generative model.
\item A high-quality generative model is chosen to generate synthetic multichannel preictal samples. The effectiveness of the model is confirmed by comparing ES prediction performance without and with data augmentation (DA) techniques. The results reveal that synthetic preictal sample augmentation can improve ES prediction performance.
\end{itemize}

The remaining content of this paper is organized as follows. Section \ref{Related} describes previous studies applying the GAN technique to the field of epilepsy study. Section \ref{Method} elaborates on the methodology of our work including generative model training, performance evaluation and implementation of ES prediction experiments without and with DA. The performance and comparison are presented in Section \ref{Result}. Sections \ref{Dis} and \ref{Conclusion} include a discussion of our results and the main conclusions, respectively.

\section{Related Work}
\label{Related}

\subsection{Literature Review}
Since proposed in 2014 \cite{goodfellow2014generative}, the GAN technique has attracted significant attention from research community owing to its success in the fields of computer vision and natural language processing. Over the past three years, numerous studies that apply the GANs to the time-series biomedical signal research are emerging. In this section, we present a concise literature review that focuses on the application of GANs on EEG signal synthesis.

\begin{figure*}[ht]
\centerline{\includegraphics[width=\textwidth]{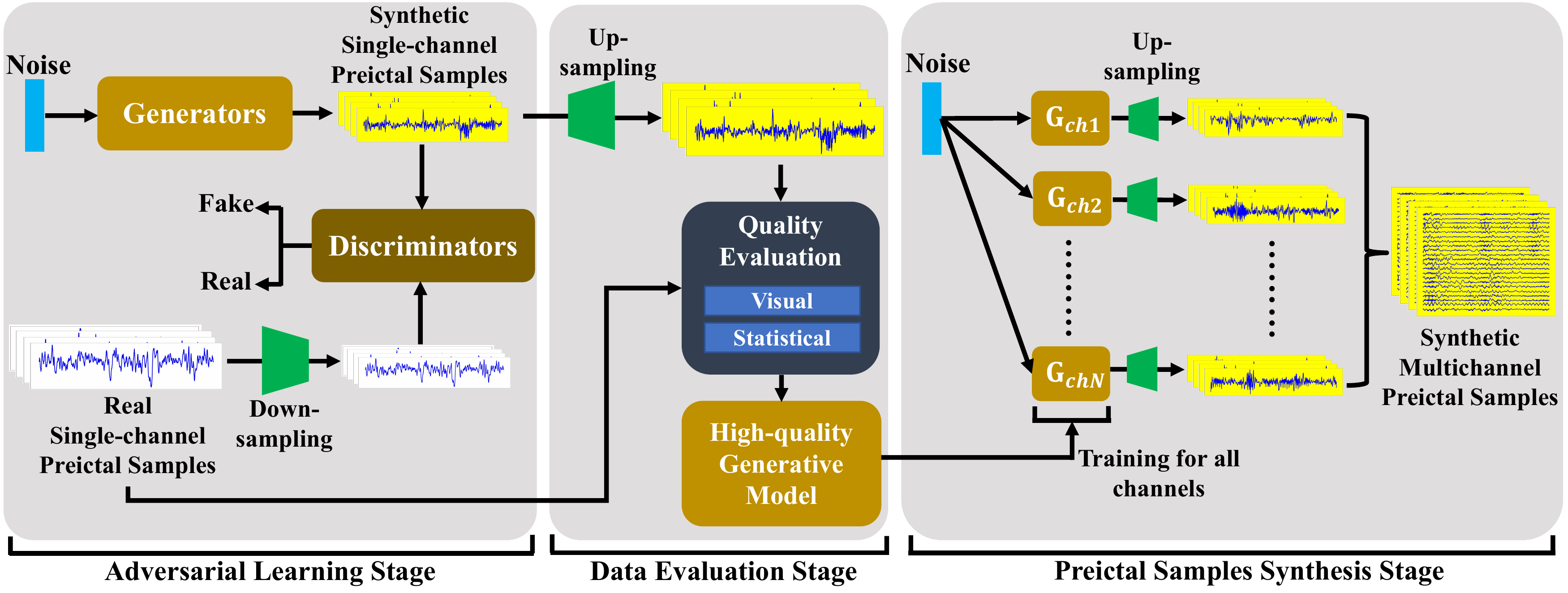}}
\caption{Flow chart of the multichannel preictal signal synthesis architecture.}
\label{fig3}
\end{figure*}

Hartmann et al. \cite{hartmann2018eeg} described an EEG-GAN framework based on deep convolution GAN (DCGAN) to synthesize naturalistic single-channel hand movement EEG signals. They compared several upsampling methods within the CNN-based generator and concluded that the cubic interpolation method outperforms other interpolation methods. Fahimi et al. \cite{fahimi2020generative} introduced a framework based on DCGAN to generat synthetic EEG signals to improve the accuracy of motor task experiment with diverted and focused attention conditions by 7.32\% and 5.45\%, respectively. Aznan et al.\cite{aznan2019simulating} utilized DCGAN, Wasserstein  GAN (WGAN), and a variational autoencoder to create synthetic EEG data for improving steady-state visual evoked potential classification tasks. They obtained slightly better performance on single-subject task, and 35\% improved accuracy on the across-subject generalization task. Luo et al. \cite{luo2018eeg} and Jiao et al. \cite{jiao2020driver} proposed a conditional WGAN to augment EEG signals to improve the classification performance in emotion recognition and driver sleepiness detection applications. In addition to CNN-based GAN, researchers \cite{abdelfattah2018augmenting,haradal2018biosignal} utilized RNN-based GAN to augment different bio-signals. Abdelfattah et al. proposed a novel GAN model and demonstrated their model can enhance classification performance on small EEG dataset. Haradal et al. confirmed their RNN-based GAN model is capable of generating synthetic bio-signals using the electrocardiogram and 
EEG datasets. Authors in \cite{9206942}, they used bidirectional RNN neurons to generate artificial motor imagery EEG signals, and Short-term Fourier Transform (STFT) and Welch’s power spectral density to evaluate the time-frequency characteristics of artificial signals.  The results indicated that GAN can capture important information of beta band. Hwang et al.\cite{JIAO2020100} proposed a scheme for zero-shot EEG signal classification using RNN-based GAN.  Their EEG encoder achieves an accuracy of 95.89\% and they reached an accuracy of 39.65\% for classification task with ten untrained EEG classes. In \cite{JIAO2020100}, authors applied RNN-based GAN model to drive sleepiness detection study based on EEG and electrooculography signals. They addressed the change in alpha waves and differentiate the two alpha-related phenomena. In addition, they adopted conditional WGAN to augment dataset and boost classifier performance.

There have also been instances of application of generative models to ES studies. Truong et al. \cite{truong2019semi} proposed a semi-supervised learning approach based on DCGAN to forecast the ESs. They took advantage of features generated from the STFT of EEG signals as real data input to generate synthetic two-dimensional feature samples. Subsequently, a discriminator was used to distinguish real and generated samples. When the training reached equilibrium, the trained discriminator was used as a feature extractor for supervised learning. However, this innovative trial did not outperform the CNN model, which had previously been proposed by the same authors for three different databases \cite{truong2018convolutional}. Rasheed et al. \cite{rasheed2020generative} proposed a generative model based on DCGAN to synthesize the spectrogram of the STFT of EEG signals for ES prediction, which utilized one-class support vector machine algorithm with radial basis function for data validation. Subsequently, the transfer learning technique was used to efficiently train the ES prediction classification model. They found that synthetic spectrogram sample augmentation increased the area under the receiver operating characteristic (ROC) curve (AUC) and sensitivity on two different databases respectively. Pascual et al. \cite{pascual2020epilepsygan} proposed a conditional GAN \cite{mirza2014conditional} approach, namely EpilepsyGAN, to synthesize 4s ictal samples from two selected channels for a seizure detection task. They compared the similarity between real and synthetic ictal samples using a spectral cosine metric, and obtained an average 1.3\% improved sensitivity for 24 patients in the ES detection task. In \cite{GAO20221}, authors took advantage of GAN as a strong candidate for data enhancement intended for tackling imbalanced EEG data distribution. After data augmentation by the GAN, they  designed one-dimensional CNN to achieve better performance on seizure detection classification task. You et al. \cite{YOU2020105472} performed automatic seizure detection with the trained GAN as an anomaly detector based on behind-the-ear EEG signals. They eventually achieved AUC of 0.939 and sensitivity of 96.3\%. In \cite{9441413}, authors presented a WGAN-based model to generate multichannel seizure data, and this approach is demonstrated to reduce the false positive rate by 72.72\%, meanwhile slightly increasing detection latency and maintaining sensitivity.

\section{Methodology of Preictal Signal Synthesis}
\label{Method}
\subsection{GAN theory}
The GAN framework consists of two opposing networks that are trained simultaneously to achieve an equilibrium. The first network, which is a generator, attempts to generate artificial realistic samples based on the latent noise input, and the second is a discriminator, which is used to distinguish between artificial and real samples. This adversarial learning aims to train the generator to capture the hidden structure and distribution of real samples to generate artificial samples with the same distribution. The loss function of GAN is defined as:

\begin{equation}
\mathop{min}\limits_{G}\mathop{max}\limits_{D}\mathop{\mathbb{E}}\limits_{\boldsymbol{x}\sim\mathbb{P}_{r}}[logD(\boldsymbol{x})] + \mathop{\mathbb{E}}\limits_{\boldsymbol{\widetilde{x}}\sim\mathbb{P}_{g}}[log(1-D(\boldsymbol{\widetilde{x}}))] 
\end{equation}

\noindent where $\mathbb{P}_{r}$ is the real data distribution and $\mathbb{P}_{g}$ represents the model distribution defined by

\begin{equation}
\widetilde{\boldsymbol{x}} = G(\boldsymbol{z})
\end{equation}

\noindent where $\boldsymbol{z}$ is a random vector from the noise distribution.

An improved GAN known as Wasserstein GAN is proposed to overcome the training stability issue of conventional GAN \cite{arjovsky2017wasserstein,gulrajani2017improved}. One drawback of the conventional GAN framework is the training instability issue in which vanishing gradients occur in the generator if the discriminator were trained to optimality during each training iteration. WGAN is intended to minimize the Wasserstein distance (WD) instead of the Jensen-Shannon divergence, which is measured by the original GAN, between real and artificial data distributions. WD, also known as the Earth-Mover distance, is used to calculate the cost of transporting mass from one distribution to another. In the WGAN, discriminator (also called critic according to \cite{arjovsky2017wasserstein}) and generator are trained to optimize the loss function:

\begin{equation}
\mathop{min}\limits_{G}\mathop{max}\limits_{D\in\mathcal{D}}\mathop{\mathbb{E}}\limits_{\boldsymbol{x}\sim\mathbb{P}_{r}}[D(\boldsymbol{x})] + \mathop{\mathbb{E}}\limits_{\boldsymbol{\widetilde{x}}\sim\mathbb{P}_{g}}[D(\boldsymbol{\widetilde{x}})] 
\end{equation}

\noindent where $\mathcal{D}$ is the set of 1-Lipschitz functions that aim to introduce Lipschitz continuity to improve the discriminator\cite{gulrajani2017improved}. For practical employment of WGAN, the weights of the discriminator are clipped to a compact interval $[-c,c]$ to enforce the Lipschitz constraint.

\subsection{Data Description}
\begin{figure}[!t]
\centerline{\includegraphics[width=6.5cm]{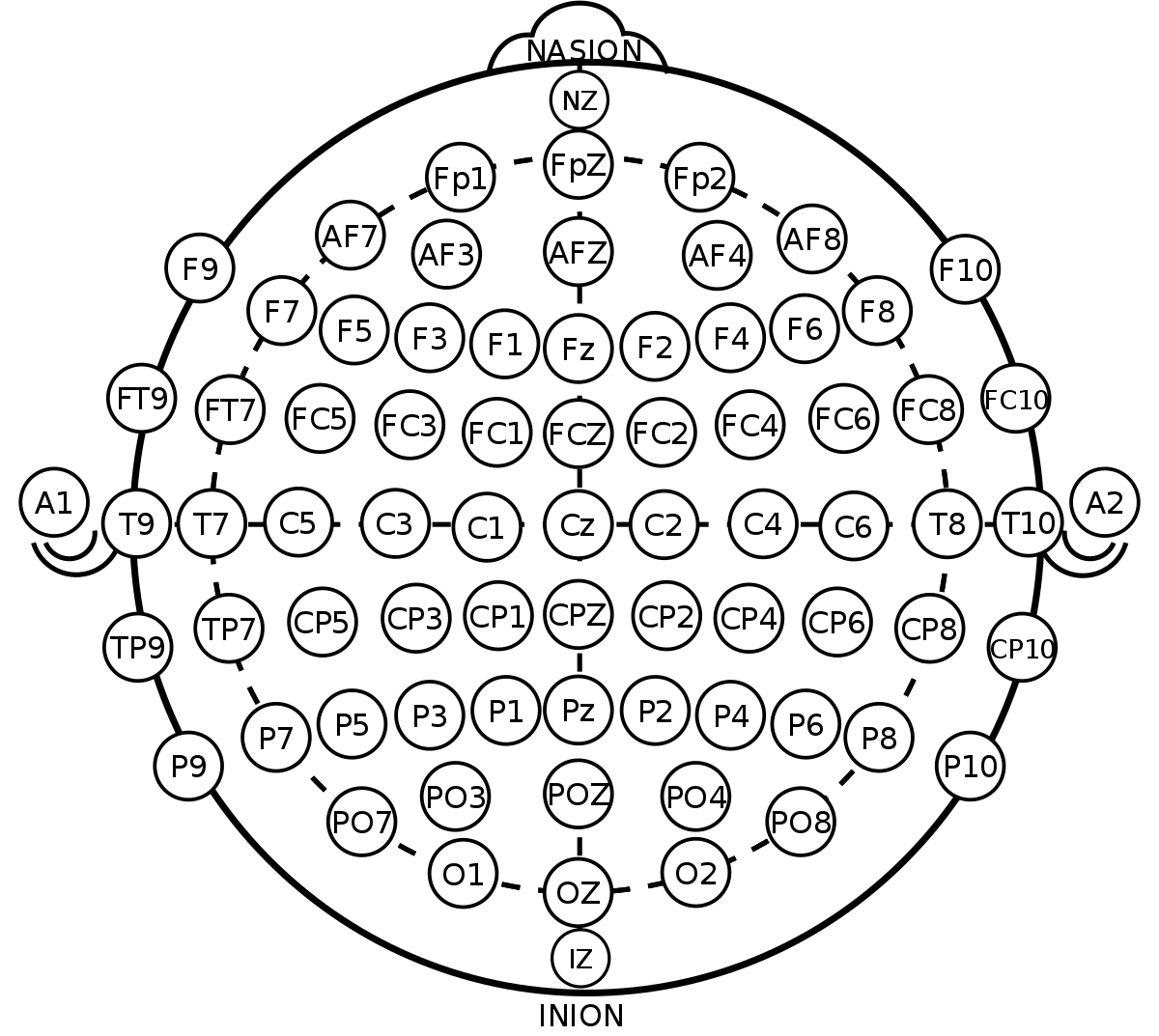}}
\caption{Electrode locations of the International 10-20 system for EEG recording \cite{klem1999ten}.}
\label{fig2}
\end{figure}

We employed and evaluated our proposed preictal signal synthesis method based on the CHB-MIT EEG dataset \cite{shoeb2009application}. The CHB-MIT dataset contains scalp EEG signals from 23 cases collected from 22 pediatric patients, and all signals were sampled at 256 Hz. The electrode position system used in this dataset was under the International 10-20 system standard. Fig. \ref{fig2} depicts the 10-20 system electrode placement \cite{klem1999ten}. However, there were electrode configuration differences among patients in terms of practical signal acquisition. We only considered patients with the same electrode configuration and no less than three leading seizures, hence 7 subjects with the same 22 electrode bipolar placement were selected for the experiments. The characteristics of selected patient subjects are shown in Table \ref{tab3} that contains the number of leading and total seizures, preictal hours and detailed electrode placement settings, where FP, F, T, P, C, and O represent frontopolar, frontal, temporal, parietal, central, and occipital respectively, odd number, even number and Z represent left side, right side and midline of the brain.

\subsection{Data Preparation}
Conventionally, preprocessing and preparation of the data is required prior to the training phase. SPH is a very short period prior to the seizure onset providing intervention, it should last a few minutes, 5min is a common choice according to previous studies \cite{daoud2019efficient,truong2018convolutional,truong2019epileptic}. Postictal period usually lasts 5-30min, we expect the postictal signal would not disturb the interictal signal, so that the longest duration 30min is chosen in this work \cite{fisher2000postictal}. The duration of an assumed preictal period varies across seizures and individuals, state-of-the-art studies usually choose the duration of preictal from 15min to 60min, and in terms of the CHB-MIT dataset, the most popular choice is 30min. Therefore, we defined the duration of the SPH, preictal and postictal periods as 5min, 30min and 30min respectively. The preictal and interictal samples are segmented from raw EEG recordings by a time-window without overlapping, the length of the time-window is chosen as 20s because the seizure prediction algorithm used in this work is based on the work published in \cite{9073988}, and the algorithm of the later work required the training and testing samples segmented in 20s. A segment of 20s was used considering much longer continuous EEG signals than 1-5s, meanwhile we considered the issues of computational cost, so that 20s is a good option. In addition, 20s duration provides flexibility for various needs of seizure prediction algorithm design because numerous state-of-the-art algorithms adopted the segments with less than 20s. The length of the single sample should be 5120 (20s $\times$ 256Hz); however, in practical experiments, we down-sampled the real preictal samples from the raw sampling rate to 100Hz intended for training convenience \cite{hussein2019human}. Thus, the length of real preictal sample for GAN training was 2000 (20s $\times$ 100Hz). Furthermore, we aimed to train the channel-wise generators independently, so that the shape of the real preictal samples was 1 $\times$ 2000.

\begin{table}[!t]
\caption{Characteristics of Selected Patient Subjects.}
\begin{center}
\begin{tabular}{ c c c c} 

\hline \hline

\makecell{Patient ID} & \makecell{No. of \\ leading seizures \\/ total seizures} &  \makecell{Preictal \\ hours} & \makecell{Electrode \\ placement}\\

\hline 
\makecell{chb01 \\ chb08 \\ chb09 \\ chb10 \\ chb11 \\ chb19 \\ chb23} & \makecell{3/7 \\ 3/5 \\ 3/4 \\ 6/7 \\ 3/3 \\3/3 \\ 3/7} &  \makecell{1.5 \\ 1.5 \\ 1.5 \\ 9 \\ 1.5 \\ 1.5 \\ 1.5} & \makecell{FP1-F7, F7-T7, T7-P7, \\ P7-O1, FP1-F3, F3-C3, \\ C3-P3, P3-O1, FP2-F4, \\ F4-C4, C4-P4, P4-O2, \\ FP2-F8, F8-T8, T8-P8, \\ P8-O2, FZ-CZ, CZ-PZ, \\ P7-T7, T7-FT9, \\FT9-FT10, FT10-T8} \\

\hline \hline

\end{tabular}

\label{tab3}
\end{center}
\end{table}

\begin{figure}[!t]
	\centering
	\subfigure[]{
	\includegraphics[width=0.95\columnwidth]{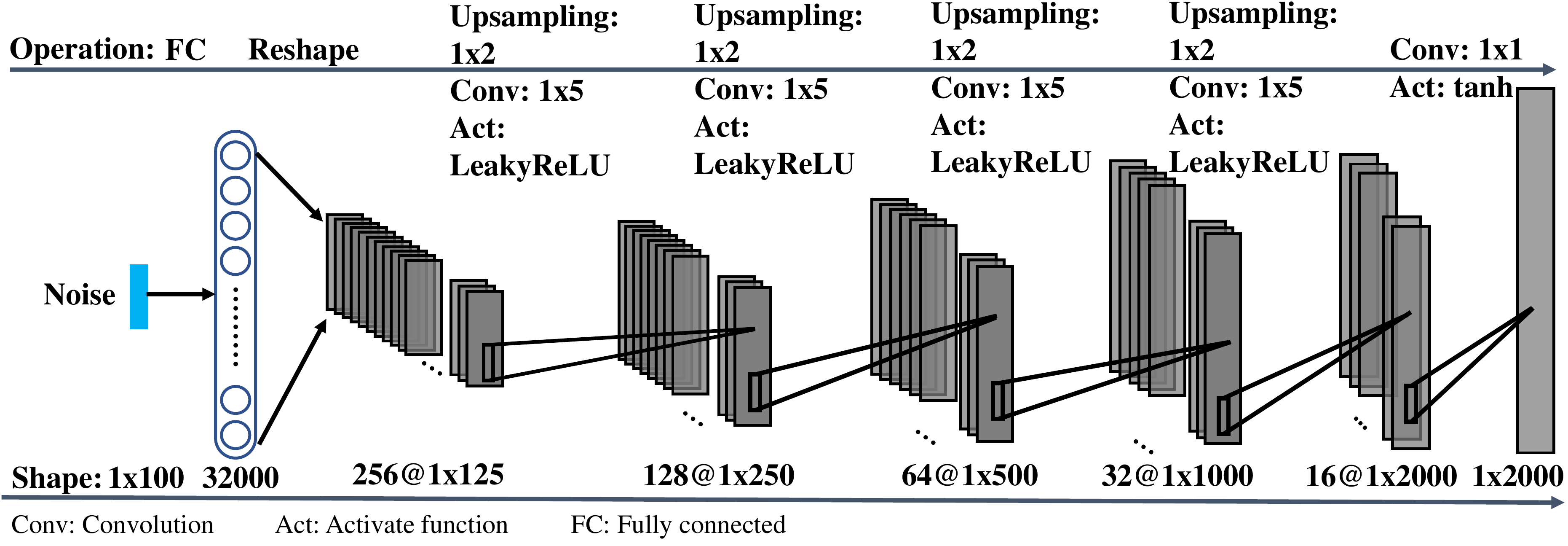}} \\
	\subfigure[]{
	\includegraphics[width=0.93\columnwidth]{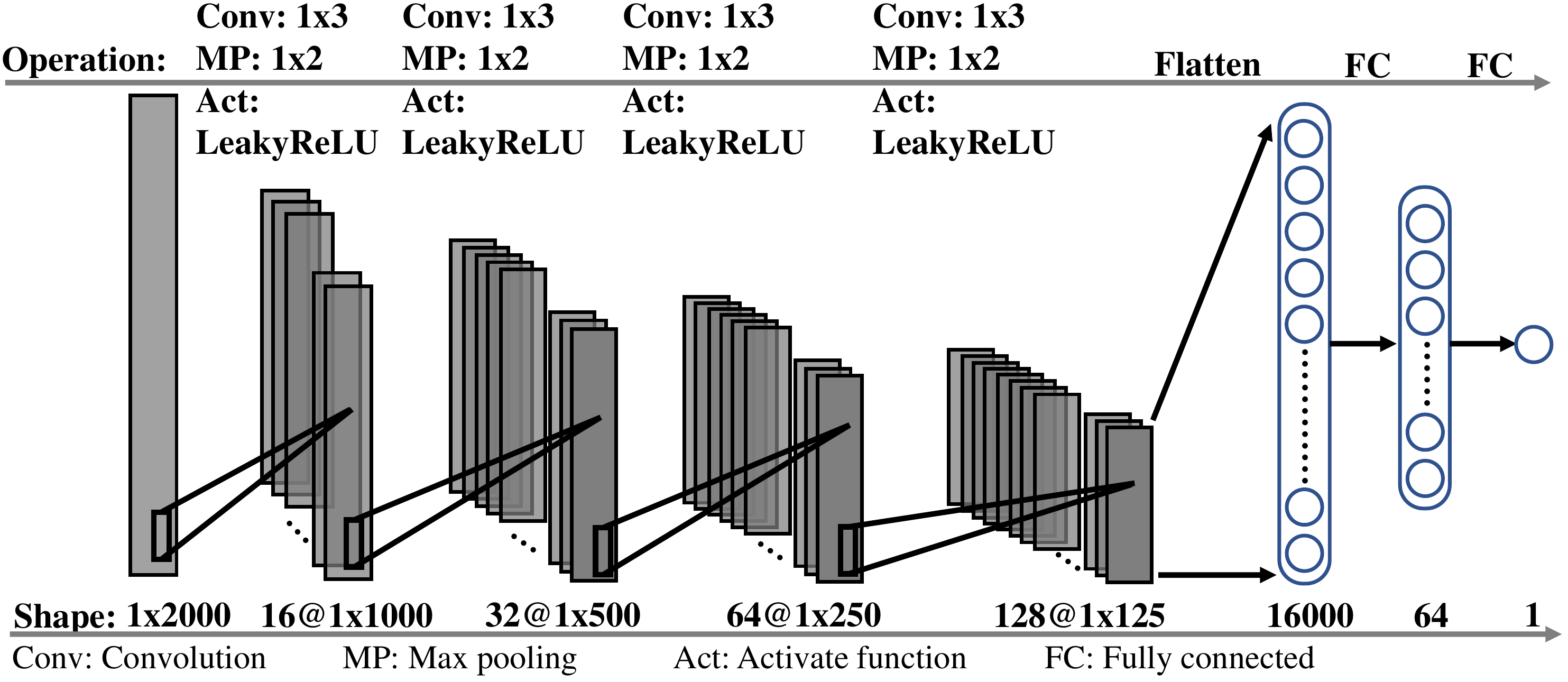}}\\

 	\caption{Convolution neural network-based architecture: \textbf{(a)} and \textbf{(b)} are the generator and discriminator, respectively.}
    
	\label{fig4}
\end{figure}

\begin{figure}[!ht]
    \centering
    \subfigure[]{\includegraphics[width=0.425\columnwidth]{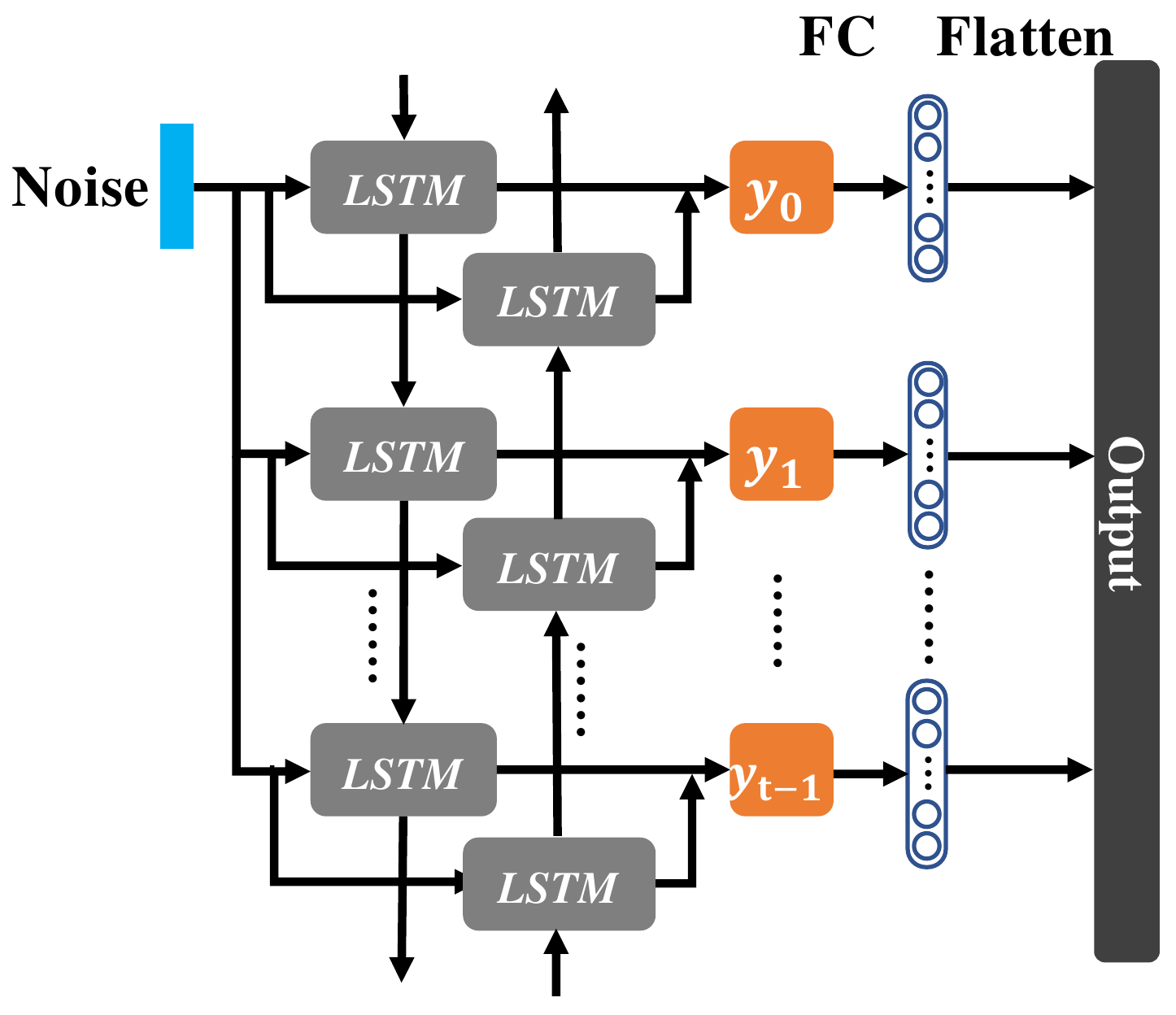}}
    \subfigure[]{\includegraphics[width=0.47\columnwidth]{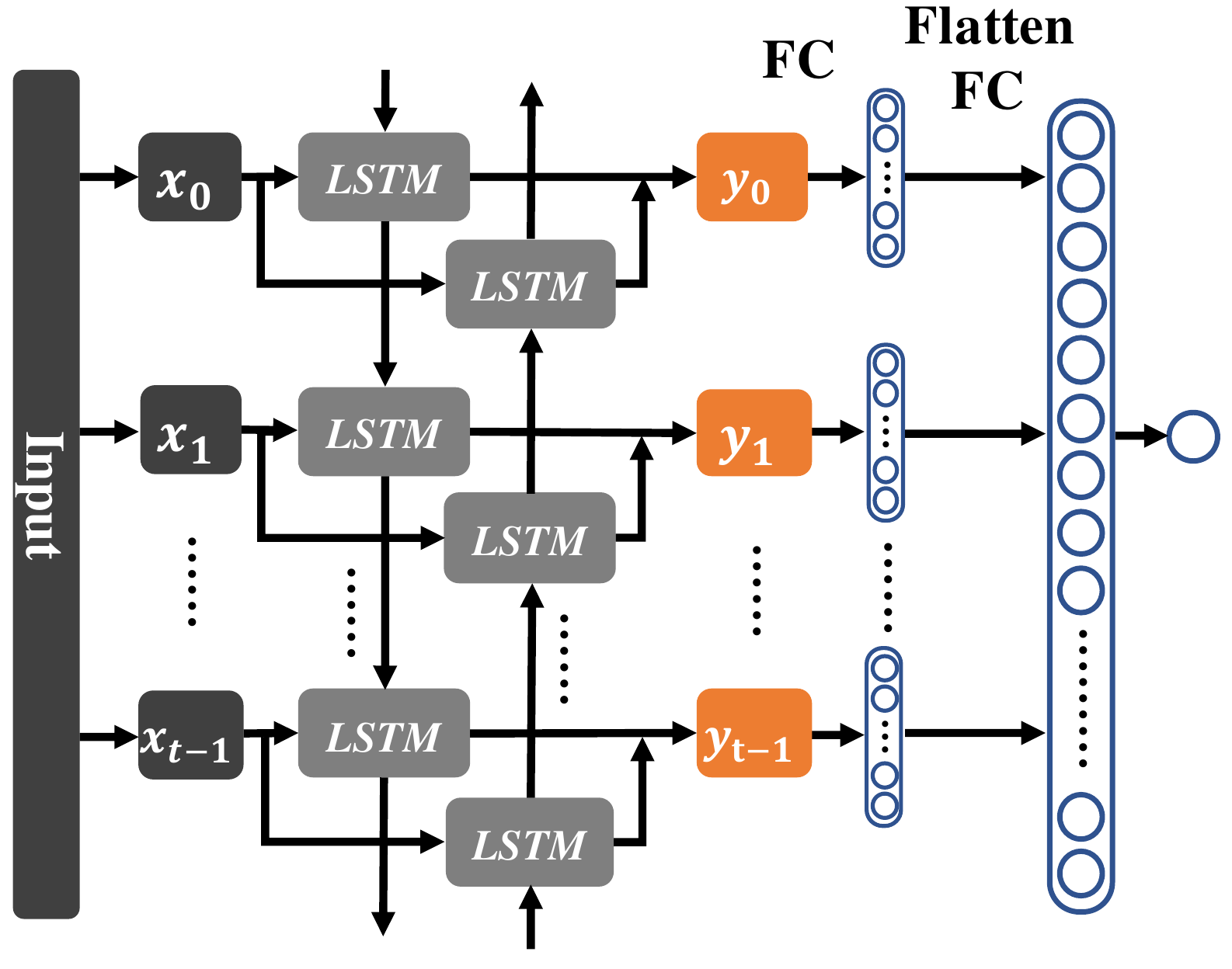}}
    \caption{Recurrent neural network-based architecture: \textbf{(a)} and \textbf{(b)} are the generator and discriminator, respectively.}
    \label{fig5}
\end{figure}

\subsection{Procedure for Preictal EEG Samples Synthesis}
Fig. \ref{fig3} shows the flow of the proposed preictal EEG sample synthesis method. It consists of three stages, i.e., adversarial learning, data evaluation and preictal sample synthesis. In the adversarial learning stage, we consider CNN- and RNN-based generative model architectures for both GAN and WGAN training. Thus, there were four candidates – DCGAN, deep convolution WGAN (DCWGAN), recurrent GAN (RGAN) and recurrent WGAN (RWGAN) for single-channel EEG sample synthesis. In the data evaluation stage, visual inspection and proposed statistical metrics were employed to evaluate the quality of synthetic single-channel samples generated by the four alternatives, and the best performing one was chosen for the latter stage. In the final stage, the high-quality generative model was used to train the channel-wise generators of all channels independently, then identical noise was added to all trained generators to synthesize signal-channel samples for all channels, the synthesized multichannel preictal EEG samples are generated by concatenating these single-channel samples. Subsequently, the effectiveness was evaluated.

\subsection{Generative Model Architecture}
The CNN- and RNN-based generator and discriminator architectures are shown in Fig. \ref{fig4} and Fig. \ref{fig5}, respectively. These figures illustrate the operation and shape of each layer; these architectures are utilized for both GAN and WGAN training. In the CNN-based GAN, the vector from the noise latent space with a shape of 1$\times$100 was fed to the generator which comprised one fully connected (FC) layer and five convolution blocks. These blocks were expected to maintain the network deep enough and the number of FC layer nodes was determined by the shape of the subsequently designed 3D tensor with a depth of 256, which was expected to contain a complex combination of patterns. Except for the last convolution block which used pixel-wise convolution to generate the output with required shape, the remaining blocks implemented the same operations in order, i.e., upsampling with the kernel shape of 1$\times$2, convolution with the kernel shape of 1$\times$5, and a leaky ReLU active function. For the CNN-based discriminator, we utilized four convolution blocks containing the same operations in order: convolution with the kernel shape of 1$\times$5, max pooling with the kernel shape of 1$\times$2, and two FC layers with 64 nodes and 1 node, respectively. 

In the RNN-based GAN, noise vectors were input to the generator which consisted of bidirectional long short-term memory (LSTM) cells with 20 time steps which were determined with the aim of processing 20s preictal samples. The output from each cell was fed to a FC layer with shared weights across time steps, and eventually a FC layer was used for generating output with the required shape \cite{hochreiter1997long}. The discriminator used an architecture similar to that of the generator, and there were only two differences that the input should be segmented with several time steps for later bidirectional LSTM cells, and a FC layer with a single neuron was added for sigmoid output generation with distinguishing objectives.

We randomly selected one channel for training different GAN architectures. When the training phase was complete, four different trained generators were used to synthesize single-channel preictal samples, and then we upsampled the synthetic preictal samples from the shape of 1$\times$2000 to the original shape of 1$\times$5120.

\subsection{Evaluation of Generative Models}
\label{evalu}
Evaluation of GAN models usually relies on the evaluation of the performance of artificial outputs generated by generator \cite{borji2019pros}. In this section, we elaborate on visual inspection and proposed statistical metrics for the evaluation of the quality of synthetic single-channel preictal samples generated by the four generators. The outperformed one is chosen to train the generators of all channels according to the evaluation approaches.

For visual inspection, we directly compared synthetic samples with real samples by vision. Although EEG signals are significantly different from images that can be easily recognized visually, we can roughly recognize the quality of EEG signals from the amplitude, frequency and period characteristics by vision. Visual inspection only provides us an intuitive quality of the synthetic EEG samples. Furthermore, reliable statistical evaluation metrics are required. 
\begin{algorithm}[!t]
 \caption{Algorithm for Multichannel Preictal Sample Synthesis using GAN.}
    \begin{algorithmic}[1]
    \renewcommand{\algorithmicrequire}{\textbf{Require:}}
    \renewcommand{\algorithmicensure}{\textbf{Initialization:}}
    \REQUIRE $c$, the clipping parameter. $m$, the batch size. $n_{d}$, the number of iterations of the discriminator per generator iteration.
    \ENSURE  $\theta_{d}$, discriminator's parameters. $\theta_{g}$, generator's parameters. $c=0.01$, $m = 32$

    \FOR {$channels = 1,2,\ldots,22$}
    \FOR {$epochs=1,2,\ldots,30000$}
	\FOR {$n_{d}=1,2,\ldots,5$}

	\STATE Sample $\{ \boldsymbol{z}^{(i)} \}^{m}_{i=1} \sim p(\boldsymbol{z})$ a batch from noise prior.
	\STATE Sample $\{ \boldsymbol{x}^{(i)} \}^{m}_{i=1} \sim \mathbb{P}_{r}$ a batch from the real single-channel preictal samples. 
	\STATE Updating the discriminator by ascending its stochastic gradient: \\ $\nabla_{\theta_{d}} \frac{1}{m}\sum\limits_{i=1}^{m} \left[log D(\boldsymbol{x}^{(i)}) + log(1-D(G(\boldsymbol{z}^{(i)})))  \right]$
	\STATE Clipping the parameters $\theta_{d} \leftarrow $ clip$(\theta_{d}, -c, c)$
    
    \ENDFOR
	\STATE Sample $\{ \boldsymbol{z}^{(i)} \}^{m}_{i=1} \sim p(\boldsymbol{z})$ a batch from noise prior.
	\STATE Updating the generator by ascending its stochastic gradient:\\$\nabla_{\theta_{g}} \frac{1}{m}\sum\limits_{i=1}^{m} log(1-D(G(\boldsymbol{z}^{(i)})))$
	
    \ENDFOR
    
    \STATE Generate synthetic single-channel preictal samples based on trained generator per channel: $G_{channels}(\boldsymbol{z})$  
    \ENDFOR
    \STATE Combination of each synthetic single-channel samples to generate  multichannel synthetic preictal samples using identical noise vector $\boldsymbol{z}$: \\ $\left[ G_{1}(\boldsymbol{z}),G_{2}(\boldsymbol{z}), \ldots, G_{22}(\boldsymbol{z})\right]^{T}$.
    \end{algorithmic} 
\end{algorithm}

Numerous statistical evaluation metrics can be used to evaluate the synthetic EEG signal. In this study, we selected three evaluation metrics for our application – frequency domain root mean square error (FDRMSE), Fréchet inception distance (FID) and WD. The root mean square error is widely applied to measure the difference between two signals. However, due to the similarity of real and synthetic samples in the time domain, meaningful information cannot be obtained because significant differences exist among different real samples in the time domain. Thus, the frequency domain information attracts our interest, and we compare the real and synthetic samples by calculating the RMSE of the frequency domain information. The FDRMSE is calculated as:

\begin{equation}
FDRMSE = \sqrt{\frac{1}{N}\sum_{n=1}^{N}(F_{[n]}-\hat{F}_{[n]})^2}
\end{equation}
    
\noindent where F and $\hat{F}$ represent the coefficients of the real and synthetic samples in the frequency domain.

FID is a common method designed to evaluate the quality of generative models and synthetic samples \cite{hartmann2018eeg,barratt2018note}. Similar to the inception score, FID utilizes a pre-trained classifier model to compare the information of real and synthetic samples representation in the embedding layer of the model. The Fréchet distance is used to calculate the distributions of two values in the embedding layer. Compared to the inception score, FID functions better because it is more robust to noise and sensitive to the quality of generative models.

WD was used as the third statistical metric for evaluation. As mentioned in Section 2.1, WD is a criterion for adversarial learning and can be used to measure the similarity between any two distributions. In this application, WD can also be regarded as sliced WD because the single-channel sample is one-dimensional.

Thus, we considered the frequency coefficients and distributions of the raw data and embedding layer instead of directly comparing the differences from the time domain. These evaluation metrics were used to select the best generative model, that was utilized to train the generators of all channels. Eventually channel-wise trained generators were used to synthesize multichannel preictal samples for following multichannel preictal sample evaluation. The pseudo-code of proposed multichannel preictal sample synthesis method is presented in Algorithm 1.
    
\begin{figure}[!t]
\centerline{\includegraphics[width=0.5\columnwidth]{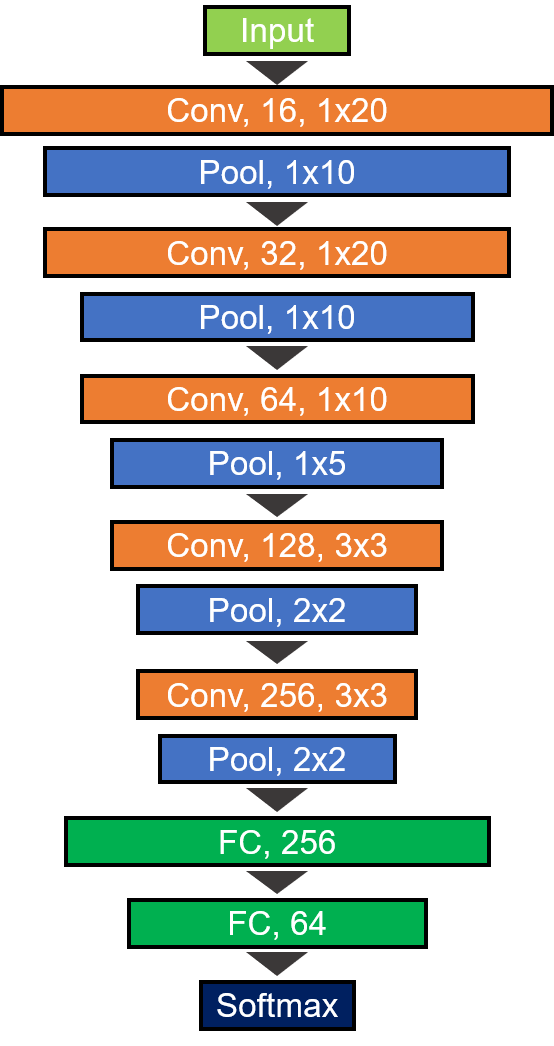}}
\caption{Architecture of previous proposed seizure prediction model. In the convolution blocks, the type of operation, the number of kernels and the shape of kernel are clarified. In terms of FC layers, the number of nodes are pointed. Conv: Convolution, FC: Fully connected layer.}
\label{PreModel}
\end{figure}    

\subsection{Evaluation of Multichannel Samples}
\begin{figure}[!ht]
	\centering
	\subfigure[]{
	\includegraphics[width=0.95\columnwidth]{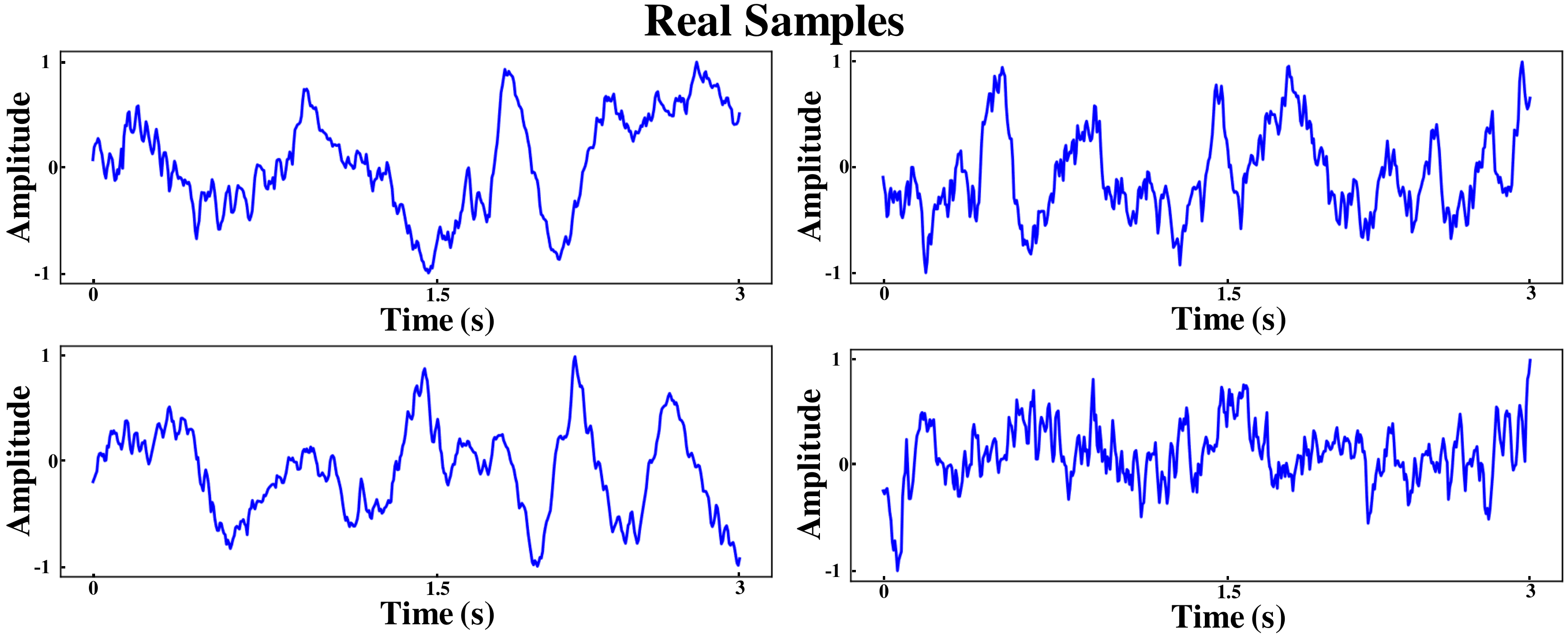}}
	\subfigure[]{
	\includegraphics[width=0.95\columnwidth]{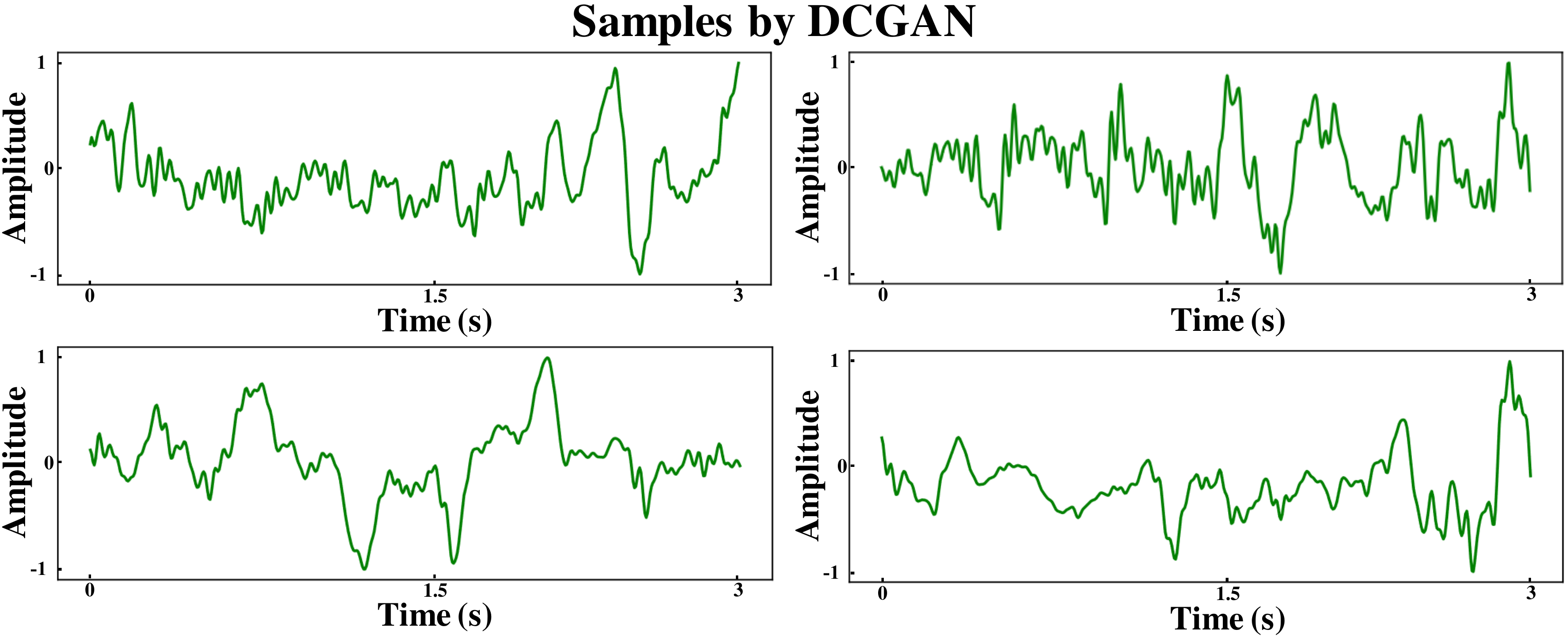}}
	\subfigure[]{
	\includegraphics[width=0.95\columnwidth]{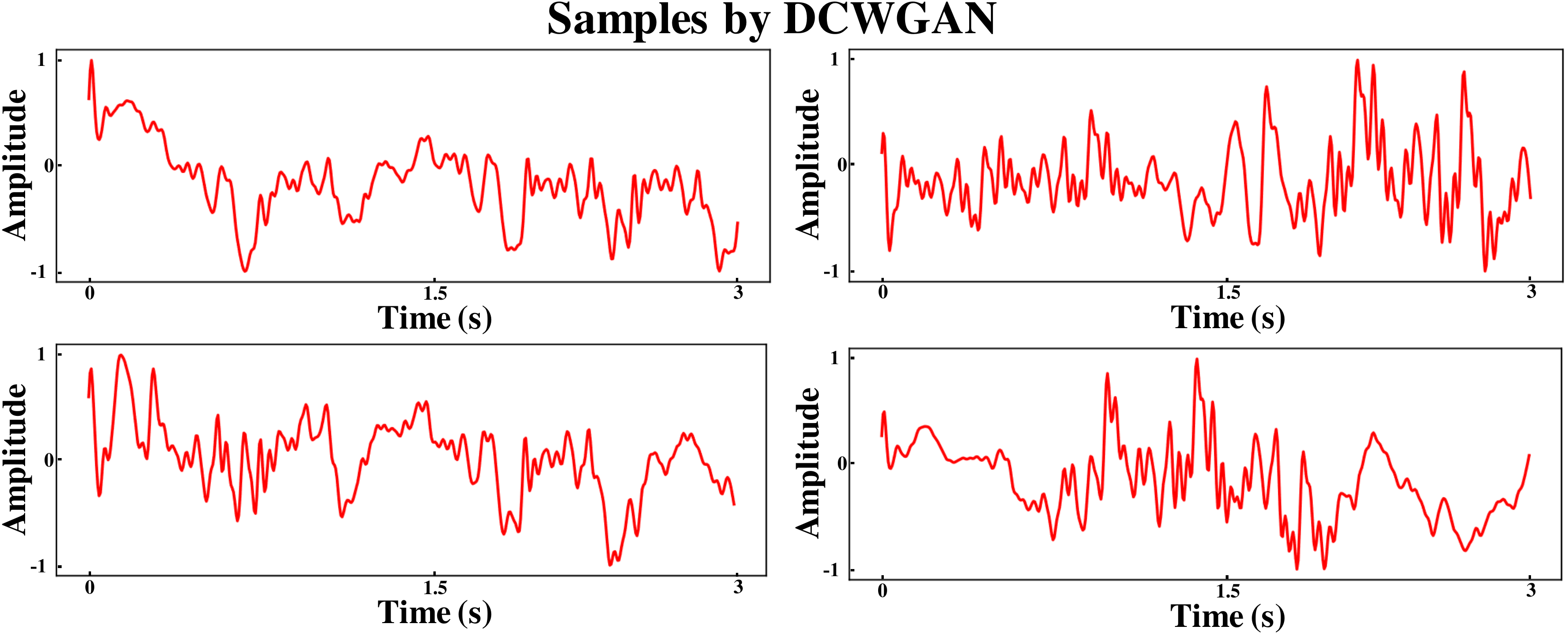}}
	\subfigure[]{
	\includegraphics[width=0.95\columnwidth]{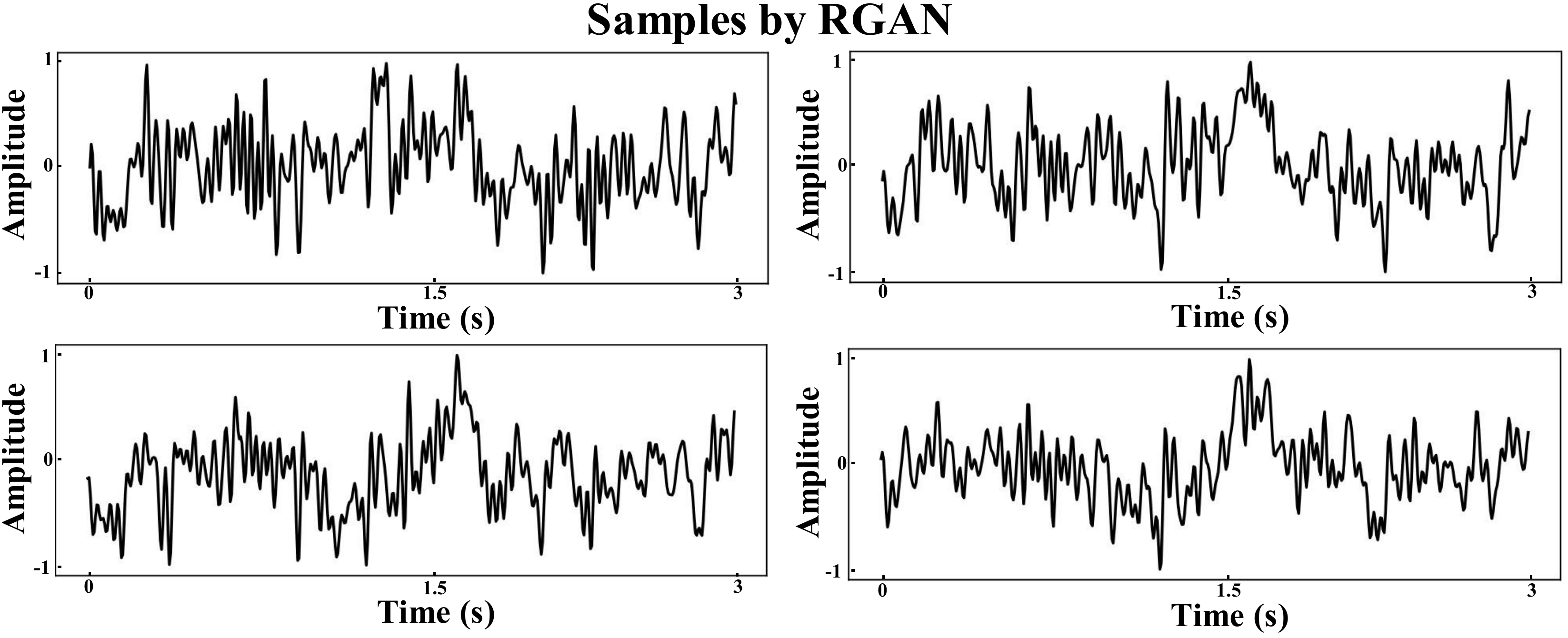}}
	\subfigure[]{
	\includegraphics[width=0.95\columnwidth]{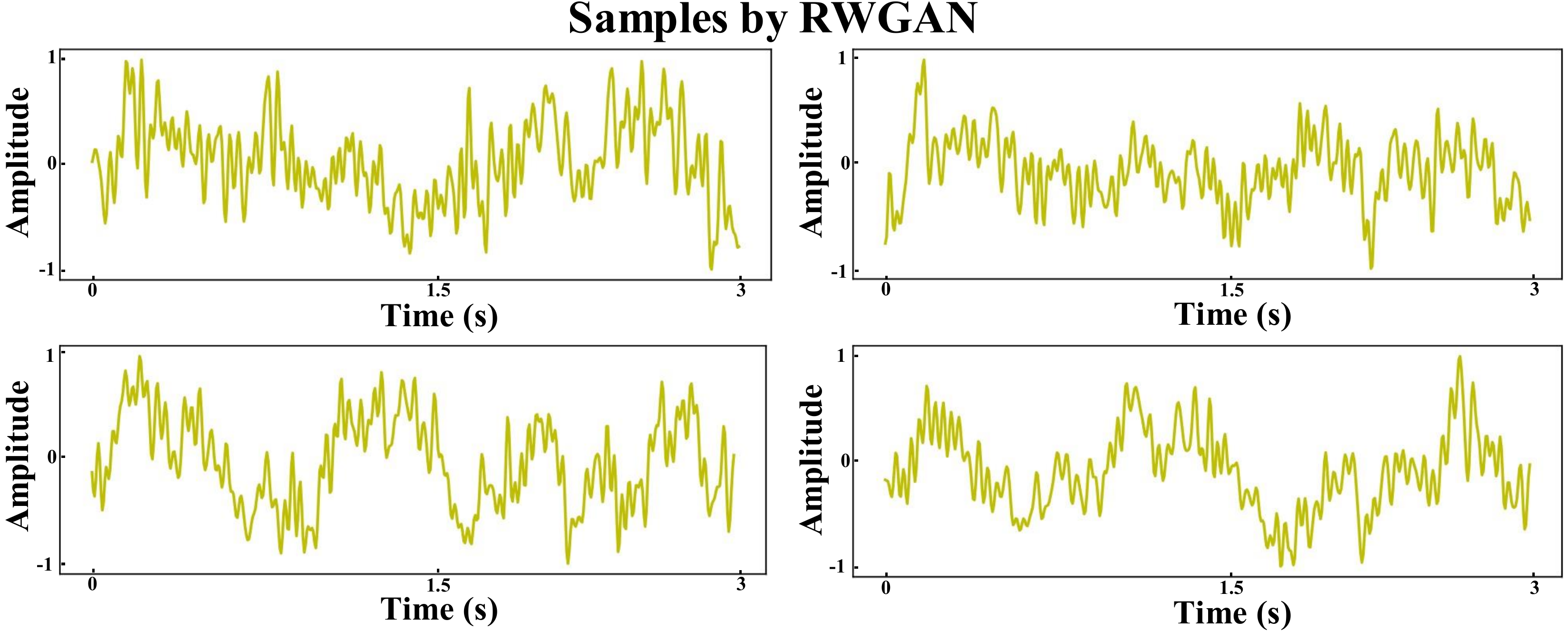}}

	\caption{Real and synthetic single-channel EEG samples. Samples with 3 s duration are shown and the amplitudes of all samples are normalized to the same range: \textbf{(a)} Real samples, \textbf{(b)}, \textbf{(c)}, \textbf{(d)} and \textbf{(e)} Synthetic samples generated by DCGAN, DCWGAN, RGAN, RWGAN, respectively. The amplitude values are scaled to [-1,1].}
    
	\label{fig6}
\end{figure}

When the outperformed generative model is chosen, we used this model to train the generators of all channels thereby synthesizing multichannel preictal samples.
After generation of a data pool containing a sufficient number of synthetic multichannel preictal samples, the effectiveness was evaluated by comparing ES prediction performance without and with synthetic samples DA. We only augmented real preictal samples with synthetic samples at different ratios, such as 3, 5 and 10 times, and the real interictal samples were kept unchanged. We utilized the classification algorithm proposed earlier by our group (Xu et al. \cite{9073988}) as a baseline model for subsequent patient-specific ES prediction experiments. Fig. \ref{PreModel} shows the architecture of the model which consists of 3 one-dimensional convolution blocks, 2 two-dimensional convolution blocks and 2 FC layers. This model aims to address raw EEG signal and learn the time-domain representations of EEG signals. We expect that the DA condition can achieve the similar or improved performance compared to the all-real condition, in order to prove the effectiveness of synthetic preictal samples.

Several statistical metrics are considered to evaluate the performance of ES prediction in this study, such as classification accuracy, ROC curve, and AUC.

\section{Results}
\label{Result}
\subsection{Generative Model Selection}
\label{hyperP}
The experiments were carried out on Python and deep learning framework Keras with Tensorflow backend, and the hardware environment is a single NVIDIA 2080Ti GPU machine. The training details are illustrated as follows. The optimizer was Adam with momentum $\beta_1=0.9$, $\beta_2=0.999$, and learning rate $\alpha=1e-5$. The leaky ReLU action function was used a slope of 0.2. A batch size of 32 and 30000 training epochs was set for training the GAN to achieve equilibrium. In addition, we trained the discriminator once and five times per generator training iteration for the conventional GAN and WGAN respectively.

\begin{table}[!t]
\caption{Evaluation scores of synthetic single-channel samples generated by different GAN models. All real and noise samples are used as a control group.}
\begin{center}
\begin{tabular}{p{1.5cm}<{\centering} | p{1.2cm}<{\centering} | p{1.2cm}<{\centering} | p{1.2cm}<{\centering} }
\hline\hline

\textbf{Model} & \textbf{FDRMSE} & \textbf{FID} & \textbf{WD}\\

\hline
DCGAN & 7.10 & 8.8 & 0.66 \\
DCWGAN & \textbf{5.39} & \textbf{7.8} & \textbf{0.45} \\
RGAN & 7.17 & 17.4 & 0.72 \\
RWGAN & 9.32 & 18.5 & 0.92 \\
\hline 
Real & 1.89 & 2.2 & 0.14 \\
Noise & 9.52 & 410 & 0.89 \\
\hline\hline
\end{tabular}

\quad
\leftline{\text{\scriptsize{\qquad \qquad  FDRMSE: Frequency domain root mean square error}}}
\leftline{\text{\scriptsize{\qquad \qquad  FID: Fréchet inception distance \quad WD: Wasserstein distance}}}
        
\label{tab1}
\end{center}
\end{table}

\begin{table}[!t]
\caption{Performance of seizure prediction experiments via leave-one-seizure-out cross validation approach. Synthetic preictal samples augmentation with three different ratios conditions are compared with all-real condition.}
\begin{center}





\begin{tabular}{ c | c | c | c | c}

\hline \hline
\makecell{Patient ID} & \makecell{All-real\\ ACC \  AUC} & \makecell{3$\times$ DA \\ ACC \ AUC} & \makecell{5$\times$ DA \\ ACC \ AUC} & \makecell{10$\times$ DA \\ ACC \ AUC} \\
\hline
\makecell{chb01 \\ chb08 \\ chb09 \\ chb10 \\ chb11 \\ chb19 \\ chb23} & \makecell{94.3  \ 0.884 \\ 86.6 \ 0.879 \\ 51.5 \ 0.449 \\ 66.7 \ 0.587 \\ 77.0 \ 0.644 \\ 44.4 \ 0.483 \\ 85.3 \ 0.809} & \makecell{95.6 \ 0.902 \\ 88.1 \ 0.881 \\ 51.5 \ 0.507 \\ 66.7 \ 0.613 \\ 77.0 \ 0.654 \\ 44.4 \ 0.501 \\ 90.3 \ 0.812} & \makecell{96.5  \ 0.910 \\ 88.5 \ 0.899 \\ 56.6 \ 0.526 \\ 69.3 \ 0.616 \\ 83.1 \ 0.663 \\ 60.7 \ 0.497 \\ 90.1 \ 0.809} & \makecell{95.9 \ 0.912 \\ 88.8 \ 0.891 \\ 56.5 \ 0.518 \\ 71.0 \ 0.637 \\ 84.7 \ 0.669 \\ 59.3 \ 0.494 \\ 90.1 \ 0.810} \\
\hline
\makecell{Average} & \makecell{73.0 \ 0.676} & \makecell{76.8 \ 0.696} & \makecell{77.8 \ 0.703} & \makecell{78.0 \ 0.704} \\

\hline \hline

\end{tabular}

\quad

\leftline{\text{\scriptsize{\quad DA: Data augmentation \quad ACC: Accuracy (\%) \quad AUC: Area under curve}}}

\label{tab2}
\end{center}
\end{table}

\begin{figure}[!t]
\centerline{\includegraphics[width=\columnwidth]{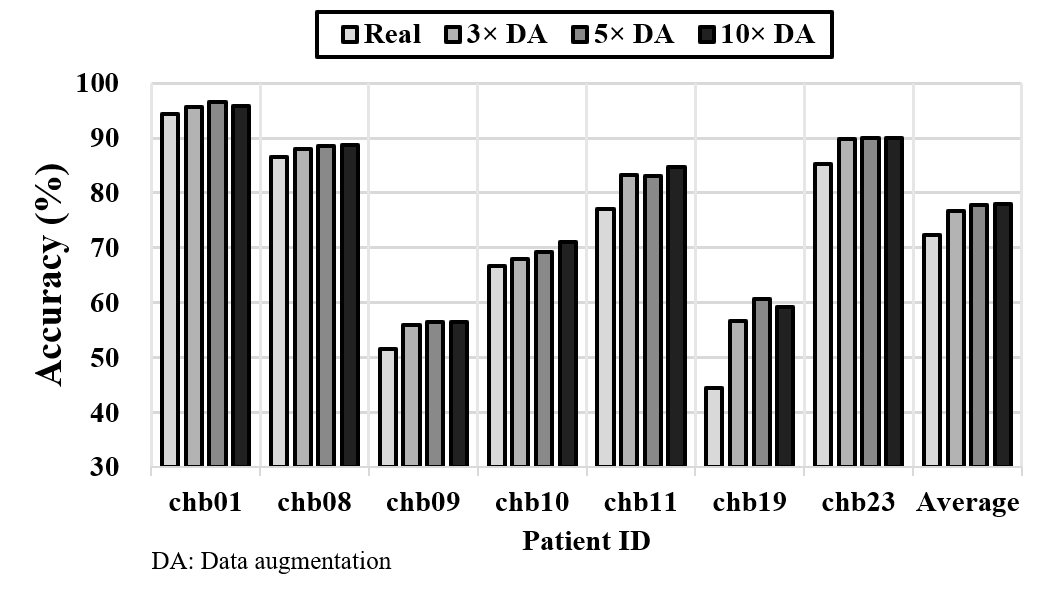}}
\caption{Accuracy histogram of seizure prediction task under the all-real and three data augmentation conditions. Average item represents averaged values of total 7 subjects.}

\label{fig7}
\end{figure}

\begin{figure}[!ht]
    \centering
    \subfigure[]{\includegraphics[width=0.49\columnwidth]{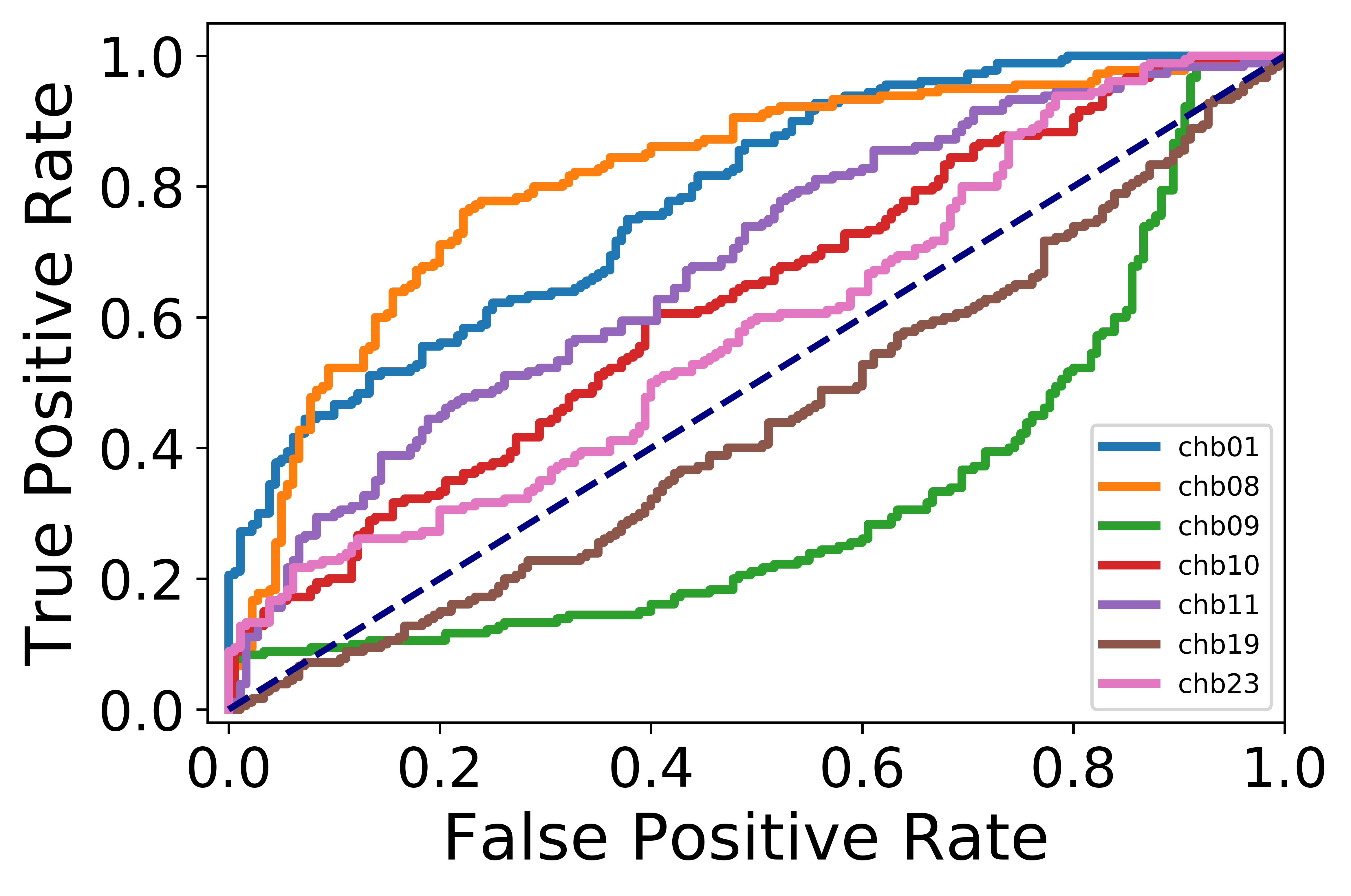}}
    \subfigure[]{\includegraphics[width=0.49\columnwidth]{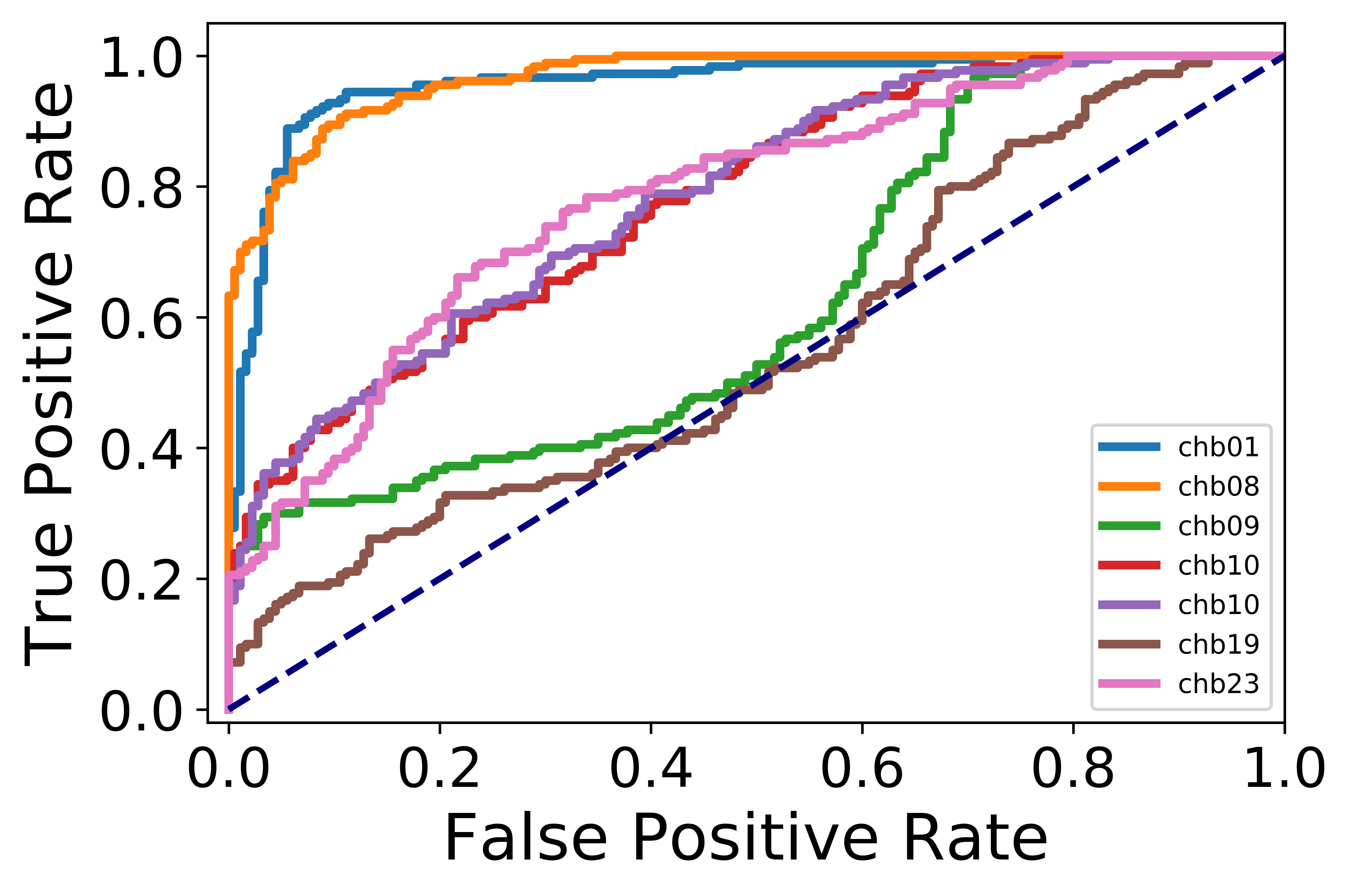}}
    \subfigure[]{\includegraphics[width=0.49\columnwidth]{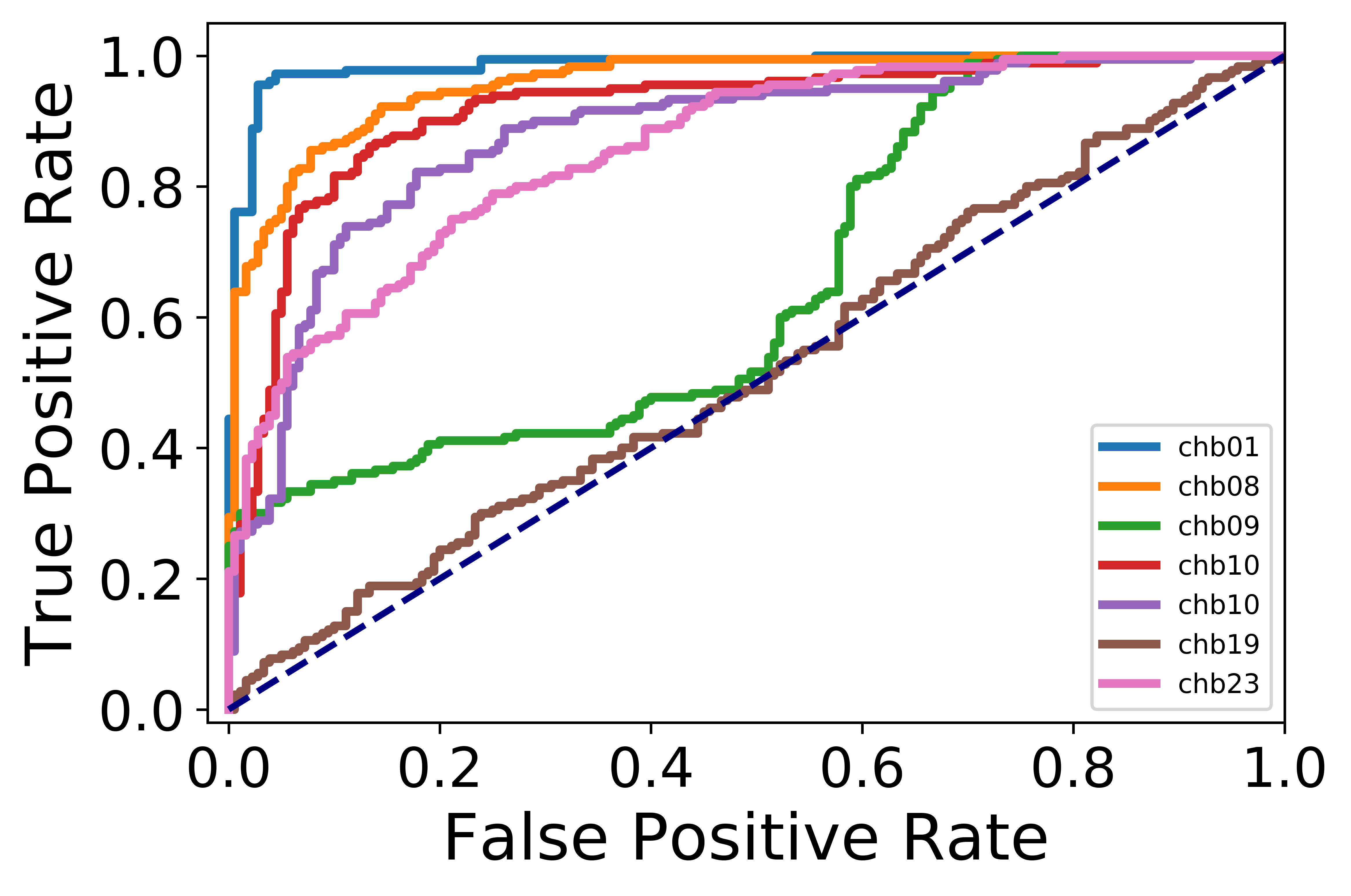}}
    \subfigure[]{\includegraphics[width=0.49\columnwidth]{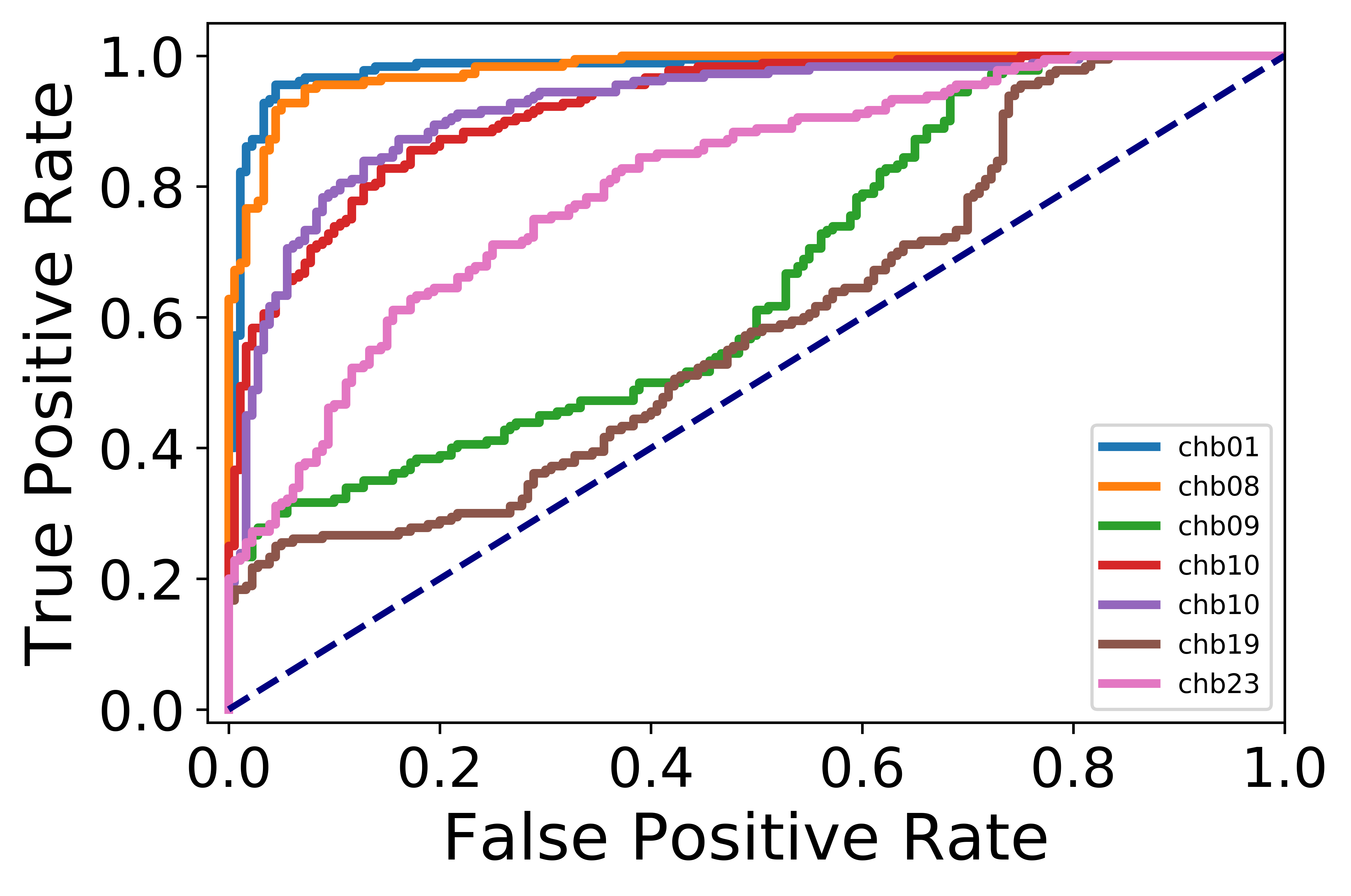}}
    \caption{ROC of seizure prediction without and with data augmentation by synthetic preictal samples. Each line represents performance of each patient-specific model respectively: \textbf{(a)} The all-real condition without data augmentation, \textbf{(b)-(d)} The data augmentation condition with 3$\times$, 5$\times$, and 10$\times$ synthetic preictal samples, respectively.}
    \label{fig8}
\end{figure}

Each trained generator could generate a data pool containing any number of synthetic preictal samples. Fig. \ref{fig6} shows the real and synthetic samples generated by the four different generators, and we display four samples randomly selected from each data pool. According to visual inspection, synthetic samples generated by DCGAN and DCWGAN are better than those generated by RGAN and RWGAN. This is because diversities in frequency, amplitude, and period are expected within the single real sample based on visual inspection, and samples should appear distinct among different real samples. However, the two RNN-based GANs both synthesize samples with similar appearances, and diversities cannot be observed within the single sample. Even though synthetic samples generated by DCGAN and DCWGAN have better appearance, we still cannot distinguish their performance by vision directly. Thus, we need statistical methods to evaluate the quality of samples generated by these two GANs.

Table \ref{tab1} shows the scores of three statistical metrics achieved by the four GANs. We randomly selected 100 samples from the data pool generated by each trained generator for score calculation, and \textit{real} and \textit{noise} items were added as a control group for comparison. We compared two different sample sets where one is selected from real samples and the other one is selected from samples of corresponding items. In terms of \textit{real} item, two different real sample sets, where each includes 100 real preictal samples randomly selected from real data pool, are compared. We accumulated total 100 scores, then averaged them to generate final score shown in Table \ref{tab1}. According to the statistical scores, we can make unanimous conclusion from visual inspection that DCGAN and DCWGAN outperform RGAN and RWGAN. Furthermore, between the two CNN-based GANs, DCWGAN achieves better scores on all three metrics than DCGAN. Therefore, we selected the DCWGAN as a high-quality model to train the generators of all channels. In terms of implementation of experiments, there are 5,778,066 parameters among the proposed DCWGAN model, and training each channel generator would cost $\sim$4h on a single GPU environment.

\begin{figure}[t]
	\centerline{\includegraphics[width=\columnwidth]{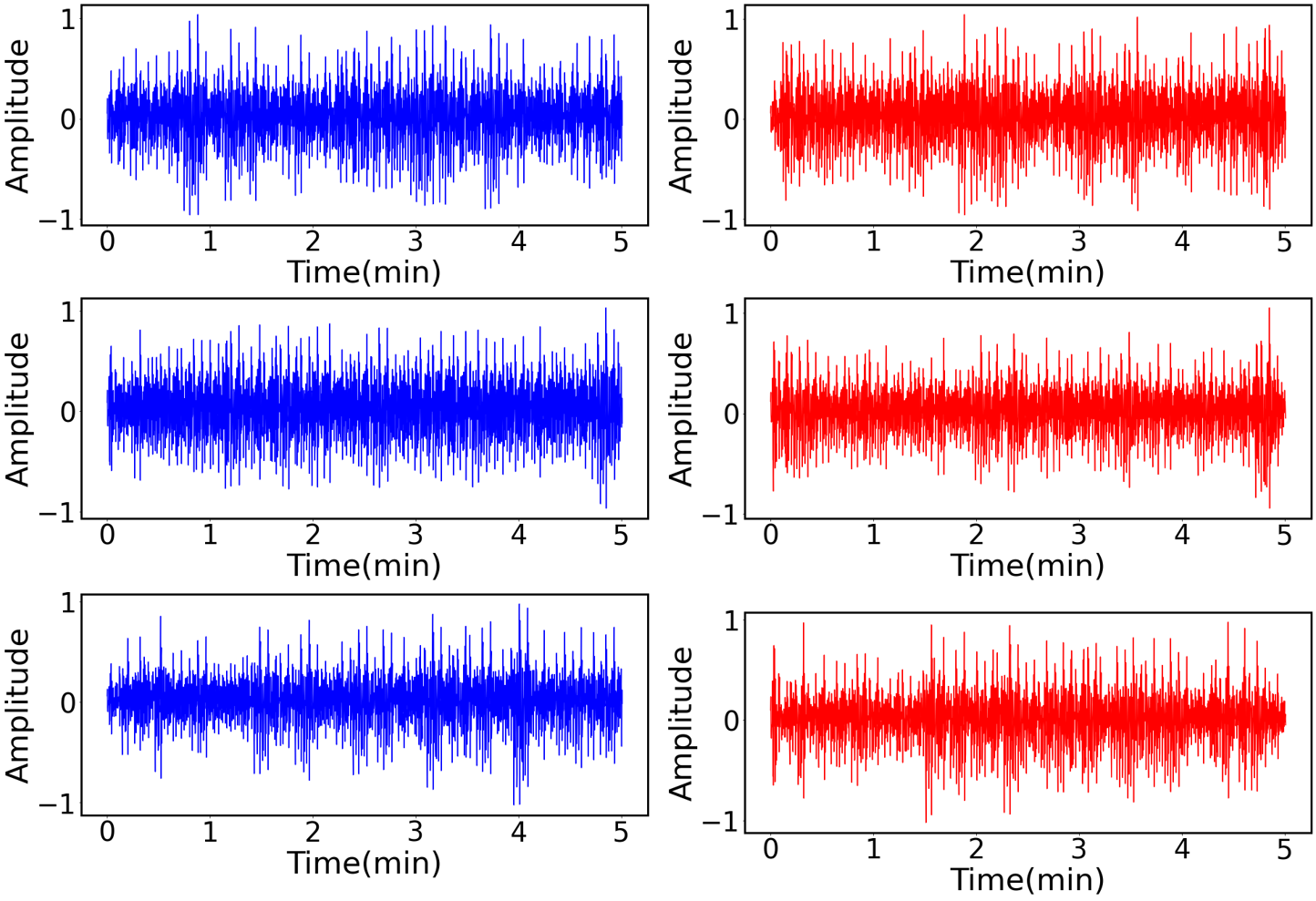}}
    
	\caption{Synthetic and real single-channel EEG recordings in 5 minutes. Blue and red signals stand for synthetic and real recordings, respectively. The amplitude values are scaled to [-1,1].}
    
	\label{fig10}
\end{figure}

\subsection{Synthetic Multichannel Preictal Samples}
\label{multiEva}
Once adversarial training of the chosen model for all channels was complete, we were able to generate synthetic complete preictal samples based on a channel-wise trained generator. With the availability of sufficient synthetic preictal samples, the effectiveness was evaluated by comparing performance of ES prediction experiments without and with augmented synthetic preictal samples. We implemented ES prediction experiments by patient-specific leave-one-seizure-out cross validation approach, preictal samples before one seizure are regarded as testing samples, and the rest preictal samples before all other seizures are used to train the model. We implemented 5 different runs and recorded averaged values of cross validation results for performance comparisons. Only preictal training set is augmented with synthetic samples at different ratios, and the number of interictal samples were always equal to the number of preictal samples for the training and testing phases. 

Table \ref{tab2} shows leave-one-seizure-out cross validation performance in the ES prediction experiments under $3\times$, $5\times$, and $10\times$ synthetic preictal sample DA conditions, all of them were used for comparison. From the overall results, all-real condition achieves accuracy of 73.0\% and AUC of 0.676, and the three DA conditions achieve accuracies of 76.8\%, 77.8\%, 78.0\% and AUC of 0.696, 0.703, 0.704, respectively. The results reveal that synthetic preictal sample augmentation can achieve improved performance $\sim$5\% in accuracy and $\sim$3\% in AUC on average compared to the results achieved from all-real data. Fig. \ref{fig7} shows the intuitively manifested histogram of the accuracy under four conditions. We can observe that data augmentation by synthetic preictal samples achieved enhanced performance for all subjects.

Fig. \ref{fig8} shows the ROC curves of patient-specific ES prediction performances without and with DA. As for each subject, we randomly selected one seizure to localize the preictal samples and randomly selected equal number of interictal samples to test the ROC performance. According to the theory of ROC curve , the line closer to the top left area can perform better. We can observe that the curves from figure (a) are distributed around the dash line, while figure (b)-(d) show that the curves are much closer to the top left area. Therefore, Fig. \ref{fig8} can also demonstrate that synthetic preictal sample augmentation can improve the performance of ES prediction.

\begin{table}[]
\caption{COMPARISON OF OUR WORK WITH PRIOR-ART PUBLICATIONS}
\label{com}
\begin{tabular}{|c|c|c|c|c|c|c|c|}
\hline
Reference                                                                        & {\cite{truong2019semi}} & {\cite{pascual2020epilepsygan}} & {\cite{YOU2020105472}} & {\cite{9441413}} & {\cite{rasheed2020generative}} & {\cite{GAO20221}} & \textbf{\begin{tabular}[c]{@{}c@{}}This\\ Work\end{tabular}} \\ \hline
Year                                                                             & 2019     & 2020     & 2020     & 2021     & 2021     & 2022     & 2022                                                         \\ \hline
\begin{tabular}[c]{@{}c@{}}Seizure\\ Prediction\\ Task?\end{tabular}             & \checkmark        & ×        & \checkmark        & ×        & \checkmark        & ×        & \checkmark                                                            \\ \hline
\begin{tabular}[c]{@{}c@{}}GAN is\\ used for \\ DA?\end{tabular}                 & ×        & \checkmark        & ×        & \checkmark        & \checkmark        & \checkmark        & \checkmark                                                            \\ \hline
\begin{tabular}[c]{@{}c@{}}Synthesize\\ Raw EEG?\end{tabular}                    & ×        & \checkmark        & ×        & \checkmark       & ×        & \checkmark        & \checkmark                                                            \\ \hline
\begin{tabular}[c]{@{}c@{}}Synthesize\\ Long-term\\ EEG\\ Signals?\end{tabular}  & ×        & ×        & ×        & \checkmark        & ×        & \checkmark        & \checkmark                                                            \\ \hline
\begin{tabular}[c]{@{}c@{}}LOSC-CV\\ Test?\end{tabular}                          & ×        & ×        & ×        & ×        & ×        & ×        & \checkmark                                                            \\ \hline
\begin{tabular}[c]{@{}c@{}}Prediction\\ Performance\\ is improved?\end{tabular} & ×        & ×        & ×        & ×        & \checkmark       & ×        & \checkmark                                                            \\ \hline
\end{tabular}

\quad

\leftline{\text{\scriptsize{LOSO-CV: Leave-One-Seizure-Out Cross Validation \quad DA: Data Augmentation}}}
\end{table}

\section{Discussion}
\label{Dis}
The results from Section \ref{multiEva} indicate augmentation with synthetic samples can be implemented effectively on either few times (3$\times$) or more times (10$\times$) condition. However, we cannot confirm that which augmentation ratio can bring best performance according to Table \ref{tab2}, because performances are similarly enhanced among three ratios. This is sufficient to demonstrate the goal of enhancing ES prediction by improved prediction performance metrics and addressing the unbalanced dataset issue to achieve robust results. Furthermore, the privacy of patient is an important issue to be considered when medical data are shared, hospitals cannot share private medical data without the permission of patients. Our proposed method that synthesis of effective artificial EEG signals is a solution enabling privacy preservation\cite{van2014systematic,clifton2004privacy}.

Most state-of-the-art seizure prediction machine learning algorithms extracted $<$30s segments from raw recordings as training samples, however, numerous recent studies raised that long-term EEG recordings in several minutes take an important role in overcoming limitations of epilepsy studies\cite{duun2020new,xinghua2020clinical}. Thus, we tested our proposed method to synthesize single-channel EEG recordings in 5 minutes. Because there is a scarce of existing algorithm using recordings in several minutes as training samples, synthetic recordings will be evaluated only by visual inspection and statistical methods. Fig. \ref{fig10} shows 3 randomly selected 5min synthetic and real recordings, we can see that they are quite similar visually. In addition, we evaluated them by 3 statistical methods proposed in Section \ref{evalu}, FDRMSE of 6.58, FID of 8.2 and WD of 0.49 are achieved between synthetic and real recordings. Thus, this can demonstrate similarity between synthetic and real recordings.

In Table \ref{com}, we compared our work with prior-art publications related to the application GAN to epilepsy studies. We can see the merits and advantages of our work according to the table.

Researchers would like to extract various features from raw EEG signals to train ES prediction classification models, few studies have processed raw EEG samples directly. In addition, researchers may need to determine features or raw signals extracted by different time windows, so that the synthesis of specific features and short-duration signals cannot be widely applied to various needs for ES prediction studies. While our synthetic preictal signal is more flexible owing to its raw EEG format.

Mode collapse is one of the most challenging aspects of GAN training. According to the trials of training our proposed GAN, the different noises would converge to the same synthetic single-channel samples if mode collapse occurs. In addition, the hyperparameters, and epochs of adversarial training also have a crucial impact on the successful training of effective generative models. In the practical experiments, we focused on tuning the hyperparameters and the number of epochs to effectively train generative models, eventually parameters are determined as illustrated in Section \ref{hyperP}. In terms of the proposed method, we trained channel-wise generative models to initially synthesize single-channel preictal samples rather than directly synthesize multichannel samples. This is because multichannel EEG signals containing a large number of data dimensions can easily fail to train the GAN. And only a few channels are still informative, it is proved that few channels or even 2 channels are also effective for the seizure prediction task \cite{jana2021deep,salant1998prediction,chang2012channel}. Hence, our single-channel generative model is also flexible for the condition of channel reduction.

\section{Conclusion}
\label{Conclusion}
We proposed a generative model based on DCWGAN to successfully synthesize effective multichannel EEG preictal samples for ES prediction. We initially generated synthetic signal-channel EEG samples and compared the quality for determining the best GAN architecture from four alternatives. Then, complete multichannel samples were synthesized by channel-wise generators. According to the comparisons of ES prediction performance, synthetic preictal sample augmentation can reach or improve the results based on the all-real data condition, which demonstrates the effectiveness of synthetic complete multichannel EEG preictal samples. 

To the best of our knowledge, this is the first study to apply a GAN to synthesize artificial multichannel EEG preictal samples for ES prediction. This method provides a solution for the scarcity of a sufficient number of preictal samples as well as privacy concerns. Furthermore, ES prediction is one of the most difficult biomedical applications that relies on EEG signals, and we believe that our method can be applied in further research efforts to synthesize EEG signals for other scenarios.

\section*{Acknowledgment}
Authors would like to acknowledge funding support from Westlake University, Zhejiang Key R$\&$D Program No. 2021C03002 and Bright Dream Joint Institute for Intelligent Robotics.

\par

\bibliographystyle{IEEEtran}
\bibliography{ref}

\begin{thebibliography}{10}
\providecommand{\url}[1]{#1}
\csname url@samestyle\endcsname
\providecommand{\newblock}{\relax}
\providecommand{\bibinfo}[2]{#2}
\providecommand{\BIBentrySTDinterwordspacing}{\spaceskip=0pt\relax}
\providecommand{\BIBentryALTinterwordstretchfactor}{4}
\providecommand{\BIBentryALTinterwordspacing}{\spaceskip=\fontdimen2\font plus
\BIBentryALTinterwordstretchfactor\fontdimen3\font minus
  \fontdimen4\font\relax}
\providecommand{\BIBforeignlanguage}[2]{{%
\expandafter\ifx\csname l@#1\endcsname\relax
\typeout{** WARNING: IEEEtran.bst: No hyphenation pattern has been}%
\typeout{** loaded for the language `#1'. Using the pattern for}%
\typeout{** the default language instead.}%
\else
\language=\csname l@#1\endcsname
\fi
#2}}
\providecommand{\BIBdecl}{\relax}
\BIBdecl

\bibitem{world2006neurological}
W.~H. Organization, \emph{Neurological disorders: public health
  challenges}.\hskip 1em plus 0.5em minus 0.4em\relax World Health
  Organization, 2006.

\bibitem{yang2020seizure}
J.~Yang and M.~Sawan, ``From seizure detection to smart and fully embedded
  seizure prediction engine: A review,'' \emph{IEEE Transactions on Biomedical
  Circuits and Systems}, vol.~14, no.~5, pp. 1008--1023, 2020.

\bibitem{assi2017towards}
E.~B. Assi, D.~K. Nguyen, S.~Rihana, and M.~Sawan, ``Towards accurate
  prediction of epileptic seizures: A review,'' \emph{Biomedical Signal
  Processing and Control}, vol.~34, pp. 144--157, 2017.

\bibitem{xu2021trends}
Y.~Xu, J.~Yang, and M.~Sawan, ``Trends and challenges of processing
  measurements from wearable devices intended for epileptic seizure
  prediction,'' \emph{Journal of Signal Processing Systems}, pp. 1--16, 2021.

\bibitem{rasheed2020machine}
K.~Rasheed, A.~Qayyum, J.~Qadir, S.~Sivathamboo, P.~Kwan, L.~Kuhlmann,
  T.~O'Brien, and A.~Razi, ``Machine learning for predicting epileptic seizures
  using {EEG} signals: A review,'' \emph{IEEE Reviews in Biomedical
  Engineering}, 2020.

\bibitem{kuhlmann2018seizure}
L.~Kuhlmann, K.~Lehnertz, M.~P. Richardson, B.~Schelter, and H.~P. Zaveri,
  ``Seizure prediction—ready for a new era,'' \emph{Nature Reviews
  Neurology}, vol.~14, no.~10, pp. 618--630, 2018.

\bibitem{hu2020epileptic}
D.~Hu, J.~Cao, X.~Lai, Y.~Wang, S.~Wang, and Y.~Ding, ``Epileptic state
  classification by fusing hand-crafted and deep learning {EEG} features,''
  \emph{IEEE Transactions on Circuits and Systems II: Express Briefs}, 2020.

\bibitem{1198251}
L.~Iasemidis, D.-S. Shiau, W.~Chaovalitwongse, J.~Sackellares, P.~Pardalos,
  J.~Principe, P.~Carney, A.~Prasad, B.~Veeramani, and K.~Tsakalis, ``Adaptive
  epileptic seizure prediction system,'' \emph{IEEE Transactions on Biomedical
  Engineering}, vol.~50, no.~5, pp. 616--627, 2003.

\bibitem{7365453}
K.~Fujiwara, M.~Miyajima, T.~Yamakawa, E.~Abe, Y.~Suzuki, Y.~Sawada, M.~Kano,
  T.~Maehara, K.~Ohta, T.~Sasai-Sakuma, T.~Sasano, M.~Matsuura, and
  E.~Matsushima, ``Epileptic seizure prediction based on multivariate
  statistical process control of heart rate variability features,'' \emph{IEEE
  Transactions on Biomedical Engineering}, vol.~63, no.~6, pp. 1321--1332,
  2016.

\bibitem{7501827}
H.-T. Shiao, V.~Cherkassky, J.~Lee, B.~Veber, E.~E. Patterson, B.~H. Brinkmann,
  and G.~A. Worrell, ``Svm-based system for prediction of epileptic seizures
  from ieeg signal,'' \emph{IEEE Transactions on Biomedical Engineering},
  vol.~64, no.~5, pp. 1011--1022, 2017.

\bibitem{8239676}
H.~Khan, L.~Marcuse, M.~Fields, K.~Swann, and B.~Yener, ``Focal onset seizure
  prediction using convolutional networks,'' \emph{IEEE Transactions on
  Biomedical Engineering}, vol.~65, no.~9, pp. 2109--2118, 2018.

\bibitem{zhao2020binary}
S.~Zhao, J.~Yang, Y.~Xu, and M.~Sawan, ``Binary single-dimensional
  convolutional neural network for seizure prediction,'' in \emph{2020 IEEE
  International Symposium on Circuits and Systems (ISCAS)}.\hskip 1em plus
  0.5em minus 0.4em\relax IEEE, 2020, pp. 1--5.

\bibitem{5415597}
L.~Chisci, A.~Mavino, G.~Perferi, M.~Sciandrone, C.~Anile, G.~Colicchio, and
  F.~Fuggetta, ``Real-time epileptic seizure prediction using ar models and
  support vector machines,'' \emph{IEEE Transactions on Biomedical
  Engineering}, vol.~57, no.~5, pp. 1124--1132, 2010.

\bibitem{hartmann2018eeg}
K.~G. Hartmann, R.~T. Schirrmeister, and T.~Ball, ``{EEG-GAN: Generative
  adversarial networks for electroencephalograhic (EEG) brain signals},''
  \emph{arXiv preprint arXiv:1806.01875}, 2018.

\bibitem{rasheed2020generative}
K.~Rasheed, J.~Qadir, T.~J. O'Brien, L.~Kuhlmann, and A.~Razi, ``A generative
  model to synthesize {EEG} data for epileptic seizure prediction,''
  \emph{arXiv preprint arXiv:2012.00430}, 2020.

\bibitem{truong2019epileptic}
N.~D. Truong, L.~Kuhlmann, M.~R. Bonyadi, D.~Querlioz, L.~Zhou, and O.~Kavehei,
  ``Epileptic seizure forecasting with generative adversarial networks,''
  \emph{IEEE Access}, vol.~7, pp. 143\,999--144\,009, 2019.

\bibitem{esteban2017real}
C.~Esteban, S.~L. Hyland, and G.~R{\"a}tsch, ``Real-valued (medical) time
  series generation with recurrent conditional {GANs},'' \emph{arXiv preprint
  arXiv:1706.02633}, 2017.

\bibitem{goodfellow2014generative}
I.~J. Goodfellow, J.~Pouget-Abadie, M.~Mirza, B.~Xu, D.~Warde-Farley, S.~Ozair,
  A.~Courville, and Y.~Bengio, ``Generative adversarial networks,'' \emph{arXiv
  preprint arXiv:1406.2661}, 2014.

\bibitem{fahimi2020generative}
F.~Fahimi, S.~Dosen, K.~K. Ang, N.~Mrachacz-Kersting, and C.~Guan, ``Generative
  adversarial networks-based data augmentation for brain-computer interface,''
  \emph{IEEE Transactions on Neural Networks and Learning Systems}, 2020.

\bibitem{aznan2019simulating}
N.~K.~N. Aznan, A.~Atapour-Abarghouei, S.~Bonner, J.~D. Connolly,
  N.~Al~Moubayed, and T.~P. Breckon, ``Simulating brain signals: Creating
  synthetic {EEG} data via neural-based generative models for improved ssvep
  classification,'' in \emph{2019 International Joint Conference on Neural
  Networks (IJCNN)}.\hskip 1em plus 0.5em minus 0.4em\relax IEEE, 2019, pp.
  1--8.

\bibitem{luo2018eeg}
Y.~Luo and B.-L. Lu, ``{EEG} data augmentation for emotion recognition using a
  conditional wasserstein {GAN},'' in \emph{2018 40th Annual International
  Conference of the IEEE Engineering in Medicine and Biology Society
  (EMBC)}.\hskip 1em plus 0.5em minus 0.4em\relax IEEE, 2018, pp. 2535--2538.

\bibitem{jiao2020driver}
Y.~Jiao, Y.~Deng, Y.~Luo, and B.-L. Lu, ``Driver sleepiness detection from {EEG
  and EOG} signals using {GAN and LSTM} networks,'' \emph{Neurocomputing}, vol.
  408, pp. 100--111, 2020.

\bibitem{abdelfattah2018augmenting}
S.~M. Abdelfattah, G.~M. Abdelrahman, and M.~Wang, ``Augmenting the size of
  {EEG} datasets using generative adversarial networks,'' in \emph{2018
  International Joint Conference on Neural Networks (IJCNN)}.\hskip 1em plus
  0.5em minus 0.4em\relax IEEE, 2018, pp. 1--6.

\bibitem{haradal2018biosignal}
S.~Haradal, H.~Hayashi, and S.~Uchida, ``Biosignal data augmentation based on
  generative adversarial networks,'' in \emph{2018 40th Annual International
  Conference of the IEEE Engineering in Medicine and Biology Society
  (EMBC)}.\hskip 1em plus 0.5em minus 0.4em\relax IEEE, 2018, pp. 368--371.

\bibitem{9206942}
S.~Roy, S.~Dora, K.~McCreadie, and G.~Prasad, ``Mieeg-gan: Generating
  artificial motor imagery electroencephalography signals,'' in \emph{2020
  International Joint Conference on Neural Networks (IJCNN)}, 2020, pp. 1--8.

\bibitem{JIAO2020100}
Y.~Jiao, Y.~Deng, Y.~Luo, and B.-L. Lu, ``Driver sleepiness detection from eeg
  and eog signals using gan and lstm networks,'' \emph{Neurocomputing}, vol.
  408, pp. 100--111, 2020.

\bibitem{truong2019semi}
N.~D. Truong, L.~Zhou, and O.~Kavehei, ``Semi-supervised seizure prediction
  with generative adversarial networks,'' in \emph{2019 41st Annual
  International Conference of the IEEE Engineering in Medicine and Biology
  Society (EMBC)}.\hskip 1em plus 0.5em minus 0.4em\relax IEEE, 2019, pp.
  2369--2372.

\bibitem{truong2018convolutional}
N.~D. Truong, A.~D. Nguyen, L.~Kuhlmann, M.~R. Bonyadi, J.~Yang, S.~Ippolito,
  and O.~Kavehei, ``Convolutional neural networks for seizure prediction using
  intracranial and scalp electroencephalogram,'' \emph{Neural Networks}, vol.
  105, pp. 104--111, 2018.

\bibitem{pascual2020epilepsygan}
D.~Pascual, A.~Amirshahi, A.~Aminifar, D.~Atienza, P.~Ryvlin, and
  R.~Wattenhofer, ``Epilepsygan: Synthetic epileptic brain activities with
  privacy preservation,'' \emph{IEEE Transactions on Biomedical Engineering},
  2020.

\bibitem{mirza2014conditional}
M.~Mirza and S.~Osindero, ``Conditional generative adversarial nets,''
  \emph{arXiv preprint arXiv:1411.1784}, 2014.

\bibitem{GAO20221}
B.~Gao, J.~Zhou, Y.~Yang, J.~Chi, and Q.~Yuan, ``Generative adversarial network
  and convolutional neural network-based eeg imbalanced classification model
  for seizure detection,'' \emph{Biocybernetics and Biomedical Engineering},
  vol.~42, no.~1, pp. 1--15, 2022.

\bibitem{YOU2020105472}
S.~You, B.~H. Cho, S.~Yook, J.~Y. Kim, Y.-M. Shon, D.-W. Seo, and I.~Y. Kim,
  ``Unsupervised automatic seizure detection for focal-onset seizures recorded
  with behind-the-ear eeg using an anomaly-detecting generative adversarial
  network,'' \emph{Computer Methods and Programs in Biomedicine}, vol. 193, p.
  105472, 2020.

\bibitem{9441413}
Y.~Guan, J.~Koerner, T.~A. Valiante, R.~Genov, and G.~O'Leary, ``Generative
  adversarial network-based synthetic seizure dataset augmentation,'' in
  \emph{2021 10th International IEEE/EMBS Conference on Neural Engineering
  (NER)}, 2021, pp. 797--800.

\bibitem{arjovsky2017wasserstein}
M.~Arjovsky, S.~Chintala, and L.~Bottou, ``Wasserstein generative adversarial
  networks,'' in \emph{International conference on machine learning}.\hskip 1em
  plus 0.5em minus 0.4em\relax PMLR, 2017, pp. 214--223.

\bibitem{gulrajani2017improved}
I.~Gulrajani, F.~Ahmed, M.~Arjovsky, V.~Dumoulin, and A.~Courville, ``Improved
  training of wasserstein {GANs},'' \emph{arXiv preprint arXiv:1704.00028},
  2017.

\bibitem{shoeb2009application}
A.~H. Shoeb, ``Application of machine learning to epileptic seizure onset
  detection and treatment,'' Ph.D. dissertation, Massachusetts Institute of
  Technology, 2009.

\bibitem{klem1999ten}
G.~H. Klem, ``The ten-twenty electrode system of the international federation.
  the internanional federation of clinical nenrophysiology,''
  \emph{Electroencephalogr. Clin. Neurophysiol. Suppl.}, vol.~52, pp. 3--6,
  1999.

\bibitem{daoud2019efficient}
H.~Daoud and M.~A. Bayoumi, ``Efficient epileptic seizure prediction based on
  deep learning,'' \emph{IEEE transactions on biomedical circuits and systems},
  vol.~13, no.~5, pp. 804--813, 2019.

\bibitem{fisher2000postictal}
R.~S. Fisher and S.~C. Schachter, ``The postictal state: a neglected entity in
  the management of epilepsy,'' \emph{Epilepsy \& Behavior}, vol.~1, no.~1, pp.
  52--59, 2000.

\bibitem{9073988}
Y.~Xu, J.~Yang, S.~Zhao, H.~Wu, and M.~Sawan, ``An end-to-end deep learning
  approach for epileptic seizure prediction,'' in \emph{2020 2nd IEEE
  International Conference on Artificial Intelligence Circuits and Systems
  (AICAS)}, 2020, pp. 266--270.

\bibitem{hussein2019human}
R.~Hussein, M.~O. Ahmed, R.~Ward, Z.~J. Wang, L.~Kuhlmann, and Y.~Guo, ``Human
  intracranial eeg quantitative analysis and automatic feature learning for
  epileptic seizure prediction,'' \emph{arXiv preprint arXiv:1904.03603}, 2019.

\bibitem{hochreiter1997long}
S.~Hochreiter and J.~Schmidhuber, ``Long short-term memory,'' \emph{Neural
  computation}, vol.~9, no.~8, pp. 1735--1780, 1997.

\bibitem{borji2019pros}
A.~Borji, ``Pros and cons of gan evaluation measures,'' \emph{Computer Vision
  and Image Understanding}, vol. 179, pp. 41--65, 2019.

\bibitem{barratt2018note}
S.~Barratt and R.~Sharma, ``A note on the inception score,'' \emph{arXiv
  preprint arXiv:1801.01973}, 2018.

\bibitem{van2014systematic}
W.~G. Van~Panhuis, P.~Paul, C.~Emerson, J.~Grefenstette, R.~Wilder, A.~J.
  Herbst, D.~Heymann, and D.~S. Burke, ``A systematic review of barriers to
  data sharing in public health,'' \emph{BMC public health}, vol.~14, no.~1,
  pp. 1--9, 2014.

\bibitem{clifton2004privacy}
C.~Clifton, M.~Kantarcioǧlu, A.~Doan, G.~Schadow, J.~Vaidya, A.~Elmagarmid,
  and D.~Suciu, ``Privacy-preserving data integration and sharing,'' in
  \emph{Proceedings of the 9th ACM SIGMOD workshop on Research issues in data
  mining and knowledge discovery}, 2004, pp. 19--26.

\bibitem{duun2020new}
J.~Duun-Henriksen, M.~Baud, M.~P. Richardson, M.~Cook, G.~Kouvas, J.~M.
  Heasman, D.~Friedman, J.~Peltola, I.~C. Zibrandtsen, and T.~W. Kjaer, ``A new
  era in electroencephalographic monitoring? subscalp devices for
  ultra--long-term recordings,'' \emph{Epilepsia}, vol.~61, no.~9, pp.
  1805--1817, 2020.

\bibitem{xinghua2020clinical}
T.~Xinghua, L.~Lin, F.~Qinyi, W.~Yarong, P.~Zheng, and L.~Zhenguo, ``The
  clinical value of long-term electroencephalogram (eeg) in seizure-free
  populations: implications from a cross-sectional study,'' \emph{BMC
  neurology}, vol.~20, no.~1, pp. 1--7, 2020.

\bibitem{jana2021deep}
R.~Jana and I.~Mukherjee, ``Deep learning based efficient epileptic seizure
  prediction with eeg channel optimization,'' \emph{Biomedical Signal
  Processing and Control}, vol.~68, p. 102767, 2021.

\bibitem{salant1998prediction}
Y.~Salant, I.~Gath, and O.~Henriksen, ``Prediction of epileptic seizures from
  two-channel eeg,'' \emph{Medical and Biological Engineering and Computing},
  vol.~36, no.~5, pp. 549--556, 1998.

\bibitem{chang2012channel}
N.-F. Chang, T.-C. Chen, C.-Y. Chiang, and L.-G. Chen, ``Channel selection for
  epilepsy seizure prediction method based on machine learning,'' in \emph{2012
  Annual International Conference of the IEEE Engineering in Medicine and
  Biology Society}.\hskip 1em plus 0.5em minus 0.4em\relax IEEE, 2012, pp.
  5162--5165.

\end{thebibliography}

\end{document}